\documentclass[12pt,preprint]{aastex}

\def\lea{\mathrel{<\kern-1.0em\lower0.9ex\hbox{$\sim$}}}
\def\gea{\mathrel{>\kern-1.0em\lower0.9ex\hbox{$\sim$}}}

\begin{document}


\title{Calcium II Triplet Spectroscopy of LMC Red Giants. I. Abundances 
and Velocities for a Sample of Populous Clusters}



\author{Aaron J. Grocholski}
\affil{Department of Astronomy, University of Florida, P.O. Box
112055, Gainesville, FL 32611; aaron@astro.ufl.edu}
\author{Andrew A. Cole\altaffilmark{1}}
\affil{Kapteyn Astronomical Institute, University of Groningen, P.O. Box 
800, 9700AV Groningen, the Netherlands; cole@astro.umn.edu}
\altaffiltext{1}{current address: Department of Astronomy, University of 
Minnesota, 116 Church St., S.E., Minneapolis, MN 55455}
\author{Ata Sarajedini}
\affil{Department of Astronomy, University of Florida, P.O. Box 112055, 
Gainesville, FL 32611; ata@astro.ufl.edu}
\author{Doug Geisler}
\affil{Departamento de Fisica, Universidad de Concepci\'{o}n, Casilla 
160-C, Concepci\'{o}n, Chile; dgeisler@astro-udec.cl}
\author{and\\ Verne V. Smith}
\affil{Gemini Project, National Optical Astronomy Observatory, Tucson, AZ 
85719; vsmith@noao.edu}








\begin{abstract}

Utilizing the FORS2 instrument on the Very Large Telescope, we have
obtained near infrared spectra for more than 200 stars in 28 populous LMC
clusters.  This cluster sample spans a large range of ages ($\sim$ 1$-$13
Gyr) and metallicities ($-0.3 \gea$ [Fe/H] $\gea -2.0$) and has good areal
coverage of the LMC disk.  The strong absorption lines of the Calcium II
triplet are used to derive cluster radial velocities and abundances.  We
determine mean cluster velocities to typically 1.6 km s$^{-1}$ and mean
metallicities to 0.04 dex (random error).  For eight of these clusters, we
report the first spectroscopically determined metallicities based on
individual cluster stars, and six of these eight have no published radial
velocity measurements.  Combining our data with archival HST/WFPC2
photometry, we find the newly measured cluster, NGC 1718, is one of the
most metal-poor ([Fe/H] $\sim$ $-$0.80), intermediate age ($\sim$ 2 Gyr)
inner disk clusters in the LMC.  Similar to what was found by previous
authors, this cluster sample has radial velocities consistent with that of
a single rotating disk system, with no indication that the newly reported
clusters exhibit halo kinematics.  Additionally, our findings confirm
previous results which show that the LMC lacks the metallicity gradient
typically seen in non-barred spiral galaxies, suggesting that the bar is
driving the mixing of stellar populations in the LMC.  However, in
contrast to previous work, we find that the higher metallicity clusters
($\gea -1.0$ dex) in our sample show a very tight distribution (mean
[Fe/H] = $-0.48$, $\sigma$ = 0.09), with no tail toward solar
metallicities.  The cluster distribution is similar to what has been found
for red giant stars in the bar, which indicates that the bar and the
intermediate age clusters have similar star formation histories.  This is
in good agreement with recent theoretical models that suggest the bar and
intermediate age clusters formed as a result of a close encounter with the
SMC $\sim$ 4 Gyr ago.


\end{abstract}


\keywords{Magellanic Clouds --- galaxies:star clusters --- 
stars:abundances}


\section{Introduction}

In the current paradigm of galaxy formation, it is believed that the
formation history of spiral galaxy spheroids, such as the Milky Way (MW)  
halo and bulge, may be dominated by the accretion/merger of smaller,
satellite galaxies (e.g.~Searle \& Zinn 1978, Zentner \& Bullock 2003).  
This type of galactic interaction is currently seen in the Sagittarius
Dwarf galaxy (Sgr), which is in the midst of being cannibalized by the MW.  
However, due to its location on the opposite side of the Galaxy from us
(Ibata, Gilmore, \& Irwin 1994), contamination by MW foreground stars
makes it difficult to study stellar populations in Sgr.  In contrast, both
the Large Magellanic Cloud (LMC) and Small Magellanic Cloud (SMC), two
satellite galaxies which may eventually be consumed into the MW halo,
suffer little from foreground contamination due to their direction on the
sky, which places them well out of the plane of the MW.  Additionally, the
relative proximity of these galaxies allows us to easily resolve stellar
populations in the Magellanic Clouds down below their oldest main sequence
turnoffs (MSTOs).  Thus, the LMC and SMC offer us a golden opportunity to
study the effects of dynamical interactions on the formation and evolution
of satellite galaxies; this information plays an integral part in
discovering the secrets of spiral galaxy formation.

One of the most direct ways to determine the chemical evolution (CEH) and
star formation history (SFH) of a galaxy is through the study of its star
clusters, which preserve a record of their host galaxy's chemical
abundances at the time of their formation.  The LMC star clusters continue
to play a critical role in shaping our understanding of the
age-metallicity relation of irregular galaxies. The rich star cluster
system of the LMC is also a unique resource for many experiments in
stellar and galactic astronomy, largely due to the fact that the LMC
harbors well populated clusters that occupy regions of the age-metallicity
plane that are devoid of MW clusters.  Thus, LMC clusters have been widely
studied as a test of stellar evolution models at intermediate metallicity
and age (e.g., Bertelli et al.~2003; Brocato, Castellani \& Piersimoni
1994; Ferraro et al.\ 1995) and as empirical templates of simple stellar
populations for applications to population synthesis models of unresolved
galaxies (e.g., Beasley, Hoyle \& Sharples 2002; Leonardi \& Rose 2003;  
Maraston 2005).



The LMC cluster system, however, is well known to show a puzzling age
distribution, with a handful of old ($\sim$ 13 Gyr), metal-poor globular
clusters, a number of intermediate age (1$-$3 Gyr), relatively metal-rich
populous clusters and apparently only one cluster, ESO 121-SC03 ($\sim$ 9
Gyr, hereafter ESO 121), that falls into the LMC's so-called `age gap'
(e.g.~Da Costa 1991; Geisler et al.~1997; Da Costa 2002).  We note that
the LMC bar seems to show a formation history very similar to that of the
clusters (Cole et al.~2005), while field SFHs derived from deep
color-magnitude diagrams (CMDs) suggest that stars in the LMC disk had a
constant, albeit low, star formation rate during the cluster `age-gap'
(e.g.~Holtzmann et al.~1999, Smecker-Hane et al.~2002). While the cause of
the cessation of cluster formation (the beginning of the age gap) is not
known, dynamical simulations by Bekki et al.~(2004) suggest that the
recent burst of cluster formation is linked to the first very close
encounter between the Clouds about 4 Gyr ago, which would have
induced ``dramatic gas cloud collisions", allowing the LMC to begin a new
epoch of cluster and star formation; strong tidal interactions between the
Clouds have likely sustained the enhanced cluster formation.  Bekki et
al.~(2004) also find that the close encounter between the clouds would
have been sufficient to cause the formation of the LMC bar around the time
of the new epoch of cluster formation, giving rise to the similar SFHs
seen in the cluster system and the bar.  In addition to enhancing star
formation, tidal forces can result in the infall or outflow of material,
thereby affecting the CEH of the LMC and, at the same time, leaving behind
a signature of the interaction.  Thus, accurate knowledge of the ages and
metallicities of LMC clusters is necessary to fully understand the
formation and dynamical history of this galaxy.

While age and metallicity estimates from isochrone fitting to CMDs exist
for a large number of clusters, the degeneracy between age and metallicity
make these estimates inherently uncertain in the absence of solid
metallicity measurements based on spectroscopic data.  Integrated light
has been used to measure [Fe/H] for many of these clusters, however, these
values are often problematic since the cluster light can be dominated by a
few luminous stars, and the results are susceptible to small-number
statistical effects.  In recent years, high spectral resolution studies of
a few prominent clusters have been undertaken, yielding, for the first
time, detailed abundance estimates of a wide variety of elements,
including iron, for {\it individual stars} within these clusters (Hill
2004;  Johnson et al.~2006).  This work is highly valuable, but because of
the large investment in telescope time necessary to obtain data of
sufficiently high signal-to-noise (S/N) ratio, it has necessarily been
limited to only a few stars in a few clusters; most of these targets are
very old, leaving the vast majority of young and intermediate age clusters
unmeasured.

Moderate resolution studies are an excellent complement to high resolution
work for a couple of reasons.  First, the multi-object capability
available for many moderate resolution spectrographs makes it possible to
observe many potential cluster members in a given field.  This increases
the probability of observing true cluster members, and facilitates their
identification, even in sparse clusters.  Second, less integration time is
needed to achieve the desired S/N ratio at moderate resolution, allowing
the observation of many more targets in a given amount of time.  Thus,
with moderate resolution spectra we can observe a large number of targets
in a short period of time and thereby create an overview of a galaxy's
global metallicity distribution, both spatial and temporal.  This approach
is particularly important for the LMC since its metallicity distribution
is very broad and the intrinsic shape is not very well known.

To date, the only large-scale spectroscopic metallicity determination for
LMC clusters based on individual cluster stars has been the landmark study
by Olszewski et al.~(1991, hereafter OSSH; Suntzeff et al.~1992).  They
obtained medium-resolution spectra of red giant branch (RGB) stars in
$\sim$ 80 clusters at a wavelength of $\approx$ 8600~\AA, centered on the
very prominent triplet of calcium II (CaT) lines. Their work was motivated
by the recognition that the CaT lines were proving to be a reliable
metallicity indicator for Galactic globular clusters (e.g., Armandroff \&
Zinn 1988; Armandroff \& Da~Costa 1991).  Additionally, this spectral
feature is easily measured in distant targets and at medium resolution
since the CaT lines are extremely strong and RGB stars are near their
brightest in the near-infrared. Using the CaT, OSSH calculated
metallicities and radial velocities for 72 of their target clusters.  
Analysis of the metallicity distribution showed that the mean [Fe/H]
values for all clusters in the inner (radius $<$ 5$\degr$) and outer
(radius $>$ 5$\degr$) LMC are almost identical ($-0.29 \pm 0.2$ and $-0.42
\pm 0.2$, respectively), suggesting the presence of little if any radial
metallicity gradient, in sharp contrast to what is seen in the MW
(e.g.~Friel et al.~2002) and M33 (Tiede, Sarajedini, \& Barker 2004).  
Using radial velocities from the OSSH sample, Schommer et al.~(1992) found
that the LMC cluster system rotates as a disk, with no indication that any
of the clusters have kinematics consistent with that of a pressure
supported halo.

However, the results of OSSH present some difficulties owing to
limitations of technology at the time.  The use of a single-slit
spectrograph severely limited the number of targets observed toward each
cluster.  Additionally, the distance of the LMC paired with a 4m telescope
required that they observe the brightest stars in the clusters.  Many of
these stars are M giants, which have spectra contaminated by TiO (although
it may not be significant until spectral type M5 or later), or are carbon
stars, neither of which are suitable for using the CaT to determine
[Fe/H].  Thus, the combination of a single-slit spectrograph with a
midsized telescope made it difficult for OSSH to build up the number of
target stars necessary to differentiate between cluster members and field
stars.  Most of the resulting cluster values are based on only one or two
stars; in some cases, there are metallicity or radial velocity
discrepancies between the few stars measured, and it is unclear which of
the values to rely on.

The interpretation of the OSSH results is further complicated by
subsequent advances both in knowledge of the globular cluster metallicity
scale to which the CaT strengths are referred (Rutledge et al.\ 1997b) and
in the standard procedure used to remove gravity and temperature
dependencies from the CaT equivalent widths (Rutledge et al.\ 1997a).  It
is not a simple matter to rederive abundances from the equivalent widths
of OSSH because of the lack of homogeneous photometry for many of the
clusters; mapping the OSSH abundances to a modern abundance scale (e.g.,
that defined at the metal-poor end by Carretta \& Gratton 1997 and near
solar metallicity by Friel et al.\ 2002) is insufficient because the
transformation is nonlinear and random metallicity errors tend to be
greatly magnified (see Cole et al.\ 2005).


In an effort to produce a modern and reliable catalog of LMC cluster
metallicities, we have obtained near-infrared spectra of an average of
eight stars in each of 28 LMC clusters.  We have taken advantage of the
multiplex capability and extraordinary image quality and light-gathering
power of the European Southern Observatory's 8.2m Very Large Telescope,
and of the great strides in the interpretation and calibration of Ca~II
triplet spectroscopy made in the past 15 years to provide accurate cluster
abundances with mean random errors of 0.04 dex. Here we present our
derived cluster metallicities and radial velocities and compare these
results to previously published spectroscopic metallicities. The
metallicity distribution of several hundred non-cluster LMC field stars
will be presented in a forthcoming paper (Cole et al., in preparation).
The current paper is laid out as follows: Section 2 discusses the
observations and data processing.  In \S3 we present the derived cluster
properties and comparisons to previous works are detailed in \S4.  
Finally, in \S5 we summarize our results.

\section{Data}

\subsection{Target Selection}

\begin{figure} 
\begin{center} 
\plotone{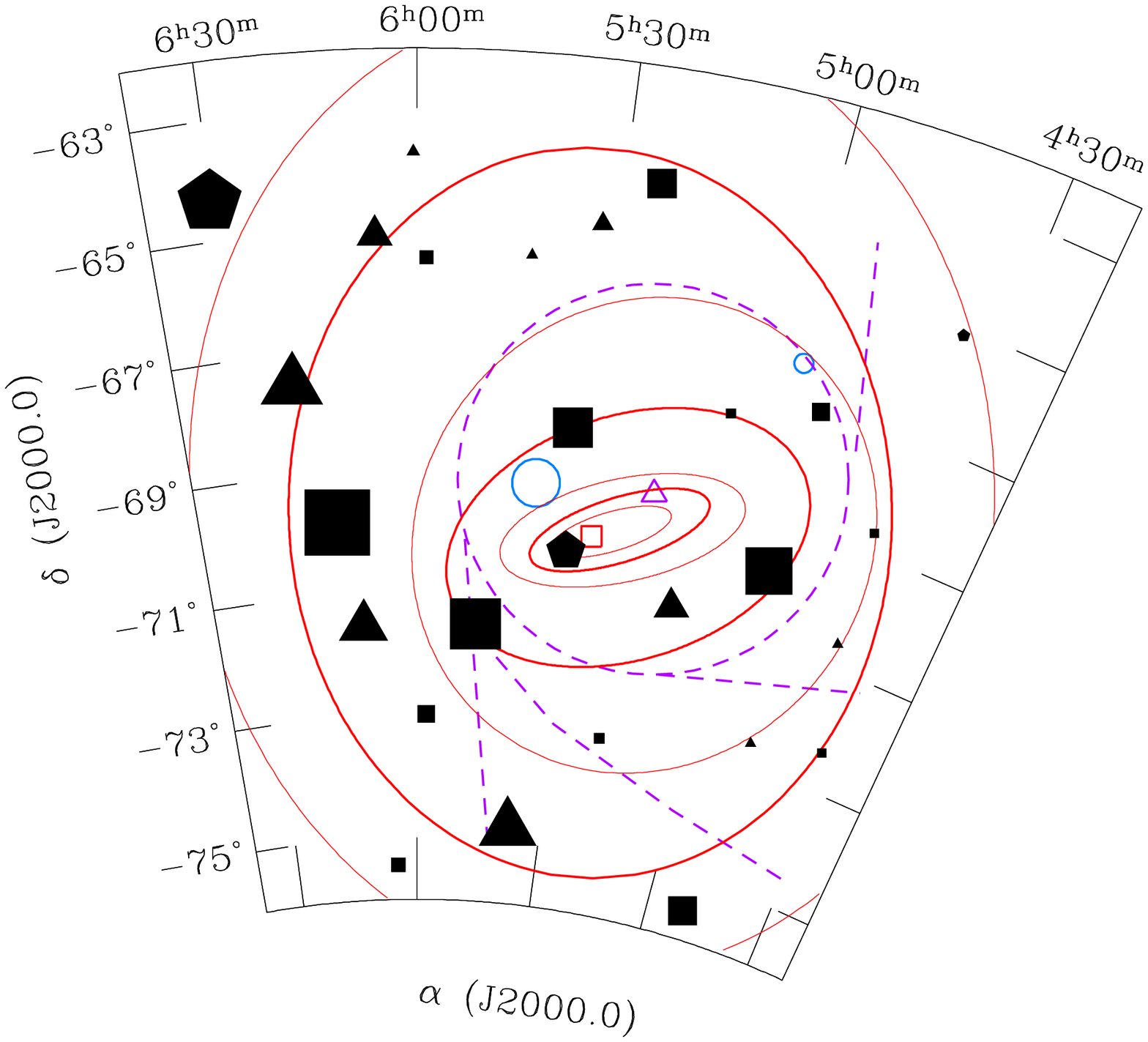}
\caption{Diagram of the LMC showing the location of our target clusters
along with prominent features in the LMC.  Filled symbols represent the
target clusters with symbol size directly related to V magnitude and shape
denoting SWB Type, where the triangles, squares and pentagons are Type V,
VI, and VII, respectively.  Note that NGC 1841 (declination $\sim
-84\degr$) is outside of the range of this plot and NGC 1861 (SWB Type
IVB) is marked as a solid triangle.  Near IR isopleths from van der Marel 
(2001) are marked as solid lines while the dashed lines outline major 
\ion{H}{1} features (see Staveley-Smith et al.~2003).  The \ion{H}{1} 
rotation center (Kim et al.~1998) is marked with the open triangle and the 
rotation center of the intermediate age stars (van der Marel et al.~2002) 
is shown by the open square.  Finally, the two largest \ion{H}{2} regions 
are marked as open circles.
}
\label{fig:schematic}
\end{center}
\end{figure}

We observed 28 prominent star clusters scattered across the face of the
LMC, in environments ranging from the dense central bar to the low-density
regions near the tidal radius (a 29th cluster was observed, however it
appears to be too young to apply the CaT method; see appendix).  Our
observations were aimed at clusters rich enough and sufficiently large and
diffuse to give us confidence in harvesting at least four definite cluster
members from which to derive the cluster metallicity.  In order to obtain
leverage on the LMC age-metallicity relation, we included clusters from
SWB class IVB$-$VII, spanning the age range of clusters containing bright,
well-populated red giant branches (Persson et al.~1983;  Ferraro et
al.~1995).  Our sample was intentionally biased towards those clusters
with conflicting or uncertain previous abundance measurements, clusters
thought to lie near the edge of the age gap, and those whose radial
velocities might provide new insight into the dynamical history of the
LMC-SMC system, based on their location.  Our targets, their positions,
sizes, integrated V magnitudes, and SWB types are listed in Table
\ref{table:info1}.  A schematic of the LMC is presented in
Fig.~\ref{fig:schematic}.  Shown are near-infrared isopleths from van der
Marel (2001; solid ellipses), at semi-major axis values of 1$\degr$,
1$\fdg$5, 2$\degr$, 3$\degr$, 4$\degr$, 6$\degr$, and 8$\degr$.  
Prominent \ion{H}{1} features (dashed lines; Staveley-Smith et al.\ 2003),
and the two largest centers of LMC star formation (30~Doradus and N11,
open circles)  are also plotted.  Finally, the rotation center of
intermediate age stars is denoted by the open square (van der Marel et
al.~2002) and the \ion{H}{1} rotation center from Kim et al.~(1998)  is
plotted as the open triangle.  Our target clusters are plotted with solid
symbols, with the exception of NGC 1841, which lies farther south than the
area covered by this diagram.

Pre-images of our target fields in $V$ and $I$ bands were taken by ESO
Paranal staff in the fall of 2004, several months prior to our observing
run.  The pre-images were processed within IRAF, and stars were identified
and photometered using the aperture photometry routines in DAOPHOT
(Stetson 1987). Stars were cataloged using the FIND routine in DAOPHOT and
photometered with an aperture size of 3 pixels. The $V$ and $I$ band data
were matched to form colors. Red giant targets were chosen based on the
instrumental CMD, and each candidate was visually inspected to ensure
location within the cluster radius (judged by eye) and freedom from
contamination by very nearby bright neighbors. In each cluster we looked
for maximum packing of the $\approx 8\arcsec$-long slits into the cluster
area and for the best possible coverage of the magnitude range from the
horizontal branch/red clump ($V \approx$ 19.2) to the tip of the RGB ($V
\approx$ 16.4).  The positions of each target were defined on the
astrometric system of the FORS2 pre-images so that the slits could be
centered as accurately as possible, and the slit identifications were
defined using the FORS Instrument Mask Simulator (FIMS) software provided
by ESO; the slit masks were cut on Paranal by the FORS2 team.

\subsection{Acquisition}
\label{sec:acquisition}

The spectroscopic observations were carried out with FORS2, in visitor
mode, at the Antu (VLT-UT1) 8.2m telescope at ESO's Paranal Observatory,
during the first half of the nights of 21-24 December 2004; weather
conditions were very clear and stable during all 4 nights, with seeing
typically 0.5$\arcsec$-1.0$\arcsec$.  We used the FORS2 spectrograph in
mask exchange unit (MXU) mode, with the 1028z+29 grism and OG590+32 order
blocking filter.  The MXU slit mask configuration allows the placement of
more slits on the sky than the 19 movable slits provided in Multi Object
Spectrograph mode.  We used slits that were 1$\arcsec$ wide and 8$\arcsec$
long (7$\arcsec$ in a few cases) and, as mentioned above, targets were
selected so as to maximize the number of likely cluster members observed;
typically 10 stars inside our estimated cluster radius were observed, with
an additional $\sim$ 20 stars outside of this radius that appeared to be
LMC field red giants, based on our preimaging CMDs.

FORS2 utilizes a pair of 2k $\times$ 4k MIT/LL CCDs and the target
clusters were centered on the upper (master) CCD, which has a readout
noise of 2.9 electrons, while the lower (secondary) CCD, with a readout
noise of 3.15 electrons, was used to observe field stars.  The only
exception to this was the Hodge 11/SL 869 field where, with a rotation of
the instrument, we were able to center Hodge 11 in the master CCD and SL
869 in the secondary CCD.  Both CCDs have an inverse gain of 0.7e$^-$
ADU$^{-1}$.  Pixels were binned 2$\times$2, yielding a plate scale of
0.25$\arcsec$ pixel$^{-1}$ and the resulting spectra cover 1750 \AA, with
a central wavelength of 8440 \AA~and a dispersion of $\sim$0.85
\AA~pixel$^{-1}$ (resolution of 2-3 \AA).  While the FORS2 field of view
is 6.8$\arcmin$ across, it is limited to 4.8$\arcmin$ of usable width in
the dispersion direction in order to keep important spectral features from
falling off the ends of the CCD.

Each field was observed twice, with offsets of 2$\arcsec$ between
exposures, to ameliorate the effects of cosmic rays, bad pixels and sky
fringing. The total exposure time in each setup was either 2x300s, 2x500s
or 2x600s. Both the readout time (26s) and setup time per field (some
6-10 minutes) were very quick and allowed us to obtain longer exposures
than originally planned in many cases.  For most of our targets with short
exposure times (300s) we combined the spectra so as to improve the S/N
ratio.  However, with the longer exposures (500s and 600s) we found that
the S/N ratio in a single exposure was adequate, and cosmic rays and bad
pixels were not a problem, so we have used only one of the pair of
exposures in our analysis.  Column 8 of Table \ref{table:info1} gives the
total exposure time that we have used in our analysis of each cluster.

Calibration exposures were taken in daytime, under the FORS2 Instrument
Team's standard calibration plan. These comprised lamp flat-field
exposures with two different illumination configurations and He-Ne-Ar lamp
exposures for each mask. Two lamp settings are required for the
flat-fields because of parasitic light in the internal FORS2 calibration
assembly.

In addition to the LMC clusters, we observed four Galactic stars clusters
(47 Tuc, M 67, NGC 2298, and NGC 288), three of which are a subsample of
the CaT calibration clusters in Cole et al.~(2004; hereafter, C04).  
Since we used the same instrument setup as C04, we expected to use their
CaT calibration, and these three clusters were observed to serve as a
check on the validity of that approach.  Processing of these three
clusters shows that our results are identical to within the errors, thus
we will use the CaT calibration of C04 rather than deriving our own CaT
calibration coefficients.

\subsection{Processing}

Image processing was performed with a variety of tasks in IRAF.  The IRAF
task CCDPROC was used to fit and subtract the overscan region, trim the
images, fix bad pixels, and flat field each image with the appropriate
dome flats.  The flat-fielded images were then corrected for distortions
in order to facilitate extraction and dispersion correction of the
spectra.  The distortion correction is a two-step process, whereby first
the image of each slitlet is rectified to a constant range of $y$-pixel
(spatial direction) values on the CCD, and then the bright sky lines are
traced along each slitlet and brought perpendicular to the dispersion
direction.  The amount of the distortion is minimal near the center of the
field of view and increases toward the edges; in all cases it is fit with
a polynomial that is at most quadratic in $y$ and linear in $x$.  
Although the distortion corrections are small, they greatly reduce the
residuals left after sky subtraction and improve the precision and
accuracy of the dispersion solution (see below).

\begin{figure}
\begin{center}
\plotone{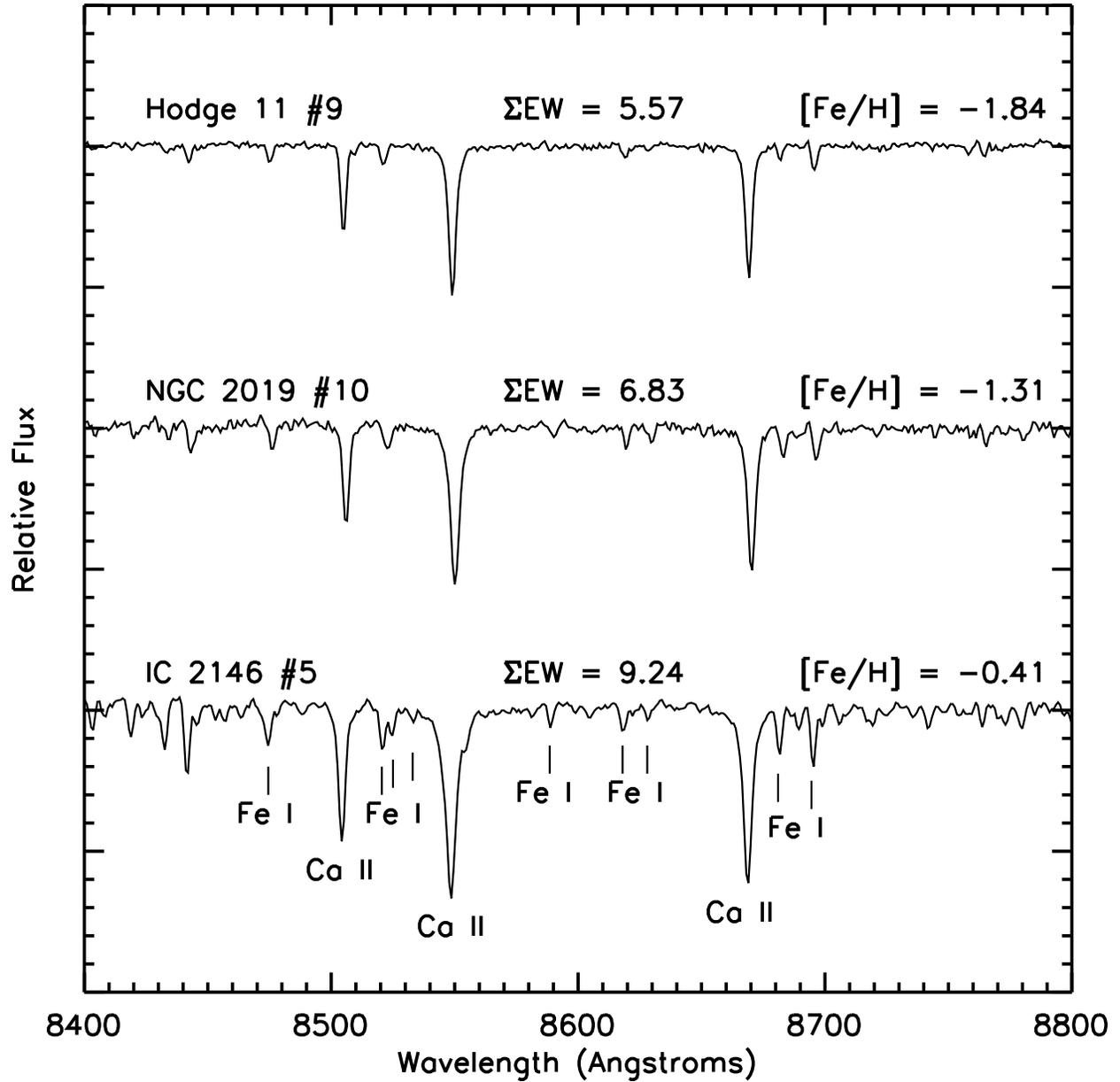}
\caption{A sample of spectra from RGB stars in our target clusters
covering a range in metallicities.  The three CaT lines, along with some
nearby \ion{Fe}{1} lines, are marked for reference; the change in CaT line
strength with [Fe/H] is readily visible.  Calculated summed equivalent 
widths and metallicities for each star are given.  
}
\label{fig:specplot}
\end{center}
\end{figure}

Once distortion corrections were completed, the task APALL (in the HYDRA
package) was used to define the sky background and extract the stellar
spectra into one-dimension.  The sky level was defined by performing a
linear fit, across the dispersion direction, to sky `windows' on each side
of the star.  This procedure presented few difficulties since the target
stars were usually bright as compared to the sky and the seeing disks are
small compared to the length of the slitlets.  The only problems arose
when the star fell near the top or bottom of the slitlet; in these cases
the sky regions were chosen interactively and we found, for all of these
spectra, that the resulting sky subtraction was indistinguishable from
that of more centrally located stars.  While daily arc lamp exposures are
available for dispersion correcting the spectra, telescope flexure during
the night, along with small slit centering errors, makes this a less
desirable method for correcting the spectra.  As such, greater than 30 OH
night sky emission lines (Osterbrock \& Martel 1992) were used by the IRAF
tasks IDENTIFY, REFSPECTRA, and DISPCOR to calculate and apply the
dispersion solution for each spectrum, which was found to be $\sim$ 0.85
\AA\ pixel$^{-1}$ with a characteristic rms scatter of $\sim$ 0.06 \AA.  
For the short (300s) exposure data, we processed both sets of images for
each pointing and combined the dispersion corrected spectra using SCOMBINE
to improve the S/N ratios for these stars.  In a few cases we found that
averaging the stellar spectra actually decreased the S/N ratio; for these
stars we chose to use the higher quality of the two individual spectra in
place of the averaged spectrum.  All spectra were then continuum
normalized by fitting a polynomial to the stellar continuum, excluding
strong absorption features (both telluric and stellar).  For the final
spectra, S/N ratios are typically 25$-$50 pixel$^{-1}$ with some stars as
high as $\sim$ 90 pixel$^{-1}$ and, in only a few cases, as low as $\sim$
15 pixel$^{-1}$.  Sample spectra showing the CaT region are presented in
Fig.~\ref{fig:specplot}.

\subsection{Radial Velocities}

Accurate radial velocities for our target stars are important for two
reasons.  First and foremost, since a cluster's velocity dispersion is
expected to be relatively small compared to the surrounding field and its
mean velocity quite possibly distinct from the field, radial velocities
are an excellent tool for determining cluster membership.  In addition,
our equivalent width measuring program uses radial velocities to derive
the expected CaT line centers.

Radial velocities for all target stars were determined through
cross-correlation with 30 template stars using the IRAF task FXCOR (Tonry
\& Davis 1979) and we have chosen to use template spectra from C04.  The
template stars were observed as a part of their CaT calibration program,
thus their sample offers a good match to the spectral types of our target
stars.  Additionally, their observations were made with a telescope and
instrument setup that is almost identical to ours.  C04 chose template
stars for which reliable published radial velocity measurements were
available.  Template velocities come from the following sources: 11 stars
from NGC 2298, NGC 1904, and NGC 4590 (Geisler et al.~1995), eight stars
from Berkeley 20 and Berkeley 39 (Friel et al.~2002), two stars from
Melotte 66 (Friel \& Janes 1993), six stars from M67 (Mathieu et
al.~1986), and three stars from 47 Tuc (Mayor et al.~1983).  In addition
to calculating relative radial velocities, FXCOR uses information about
the observatory location and the date and time of the observations (once
the ESO header has been appropriately reformatted) to correct the derived
velocities to the heliocentric reference frame.  For a star's final
heliocentric radial velocity, we adopt the average value of each
cross-correlation result.  We find good agreement amongst the template
derived velocities, with a typical standard deviation of $\sim 6$ km
s$^{-1}$ for each star.

When the stellar image is significantly smaller than the slit width,
systematic errors due to imprecise alignment of the slit center and the
stellar centroid can dominate the error budget in the radial velocity
measurements.  With the grism and CCD used here, an offset of one pixel
across the four-pixel-wide slit would introduce an error in the measured
velocity of $\approx 30$ km~s$^{-1}$.  We follow the approach of Tolstoy
et al.\ (2001) in applying a correction to each measured radial velocity
based on the individual slit offsets; following C04, we measure the slit
offsets using acquisition (so-called ``through-slit'') images taken
immediately prior to the spectroscopic measurement, and estimate a
precision of $\approx$0.14 pixels on the measured offset value.  This
introduces an error of $\pm$ 4.2 km s$^{-1}$ and, added in quadrature with
the error resulting from the velocity cross-correlations, gives an error
of roughly 7.5 km s$^{-1}$.  We adopt this as the error in measuring the
radial velocity of an individual star.


\subsection{Equivalent Widths and Abundances}


To measure the equivalent widths of the CaT lines, we have used a
previously written fortran program (see C04 for details). However, since
this region of a star's spectrum can be contaminated by weak metal lines
and, in some cases, weak molecular bands, measuring the true equivalent
width of the CaT lines at all but the highest spectral resolutions is
impossible.  Instead, we follow the method of Armandroff \& Zinn (1988)
and define continuum bandpasses on either side of each CaT feature.  In
this wavelength range, the continuum slope of a red giant star is
virtually flat, thus, the `pseudo-continuum' for each CaT line is easily
defined by a linear fit to the mean value in each pair of continuum
windows.  The `pseudo-equivalent' width is then calculated by fitting the
sum of a Gaussian and a Lorentzian, required to have a common line center,
to each CaT line with respect to the `pseudo-continuum'.  For reference,
the rest wavelengths of the line and continuum bandpasses, as defined by
Armandroff \& Zinn (1988), are listed in Table \ref{table:bandpass}.  For
many years it has been known that even at the moderate spectral
resolutions used here, a Gaussian fit to the CaT lines is susceptible to
loss of sensitivity at high metallicity because the Gaussian fails to
accurately measure the extremely broad wings of the lines (see discussion
in Rutledge et al.\ 1997a).  We follow the procedure established in C04
and add a Lorentzian profile to the Gaussian in order to recover
sensitivity to the full range of metallicities. Errors in the equivalent
width measurements were estimated by measuring the rms scatter of the data
about the fits.

A number of previous authors have calibrated the relationship between the
strengths of the three CaT lines and stellar abundance using a variety of
methods (see Table 3 in Rutledge et al.~1997b).  In all cases, a linear
combination of the individual line strengths was used to produce the
summed equivalent width, $\Sigma W$, with weighting and inclusion of lines
(some authors dropped the weakest line, $\lambda$8498 \AA)  varying based
on the quality of their data.  Since the quality of our data is such that
all three lines are well measured, we adopt the same definition for 
$\Sigma W$ as C04,
\begin{equation}
\Sigma W \equiv EW_{8498} + EW_{8542} + EW_{8662}.
\label{eq:summed}
\end{equation}
It is well known that TiO, which has a strong absorption band beginning 
near $\lambda$8440 \AA~(e.g.~Cenarro et al.~2001), can affect the spectra 
of cool ($\sim$ M5 or later), metal rich stars.  This absorption feature, 
which depresses the `pseudo-continuum' around the CaT lines and results in 
an underestimation of the measured equivalent widths, was noted by OSSH in 
some of their LMC spectra.  During processing, we checked each spectrum 
for the appearance of this TiO absorption band and found no evidence that 
TiO had affected any of our observations.

Both theoretical (J\o rgensen, Carlsson \& Johnson 1992) and empirical
(Cenarro et al.\ 2002) studies have shown that effective temperature,
surface gravity, and metallicity all play significant roles in determining
the CaT line strengths.  However, it is well-established that for red
giants of a given metallicity, there is a linear relationship between a
star's absolute magnitude and $\Sigma W$ (Armandroff \& Da Costa 1991),
where stars farther up the RGB have larger $\Sigma W$ values.  This is
primarily due to the change in surface gravity as a star moves along the
RGB; stars near the bottom of the RGB have smaller radii, thus larger
surface gravities, which increases the H$^-$ opacity.  Since H$^-$ is the
dominant opacity source in red giants, increasing the H$^-$ opacity
depresses the `pseudo-continuum', which in turn drives down the measured
value for $\Sigma W$.  To remove the effects of luminosity on $\Sigma W$,
similar to previous authors, we define a reduced equivalent width, $W'$,
as
\begin{equation} 
W' \equiv \Sigma W + \beta(V - V_{HB}), 
\label{eq:reduced} 
\end{equation}
where the introduction of the brightness of a cluster's horizontal branch
(HB), $V_{HB}$, removes any dependence on cluster distance or reddening
(see the thorough discussion in Rutledge et al.\ 1997a).  Due to the fact
that a majority of our clusters are too young and metal rich to have a
fully formed HB, we instead adopt the median value of the core helium
burning red clump (RC) stars for these clusters (see \S3 for more
information).  Values for $\beta$ have been derived empirically by
previous authors, with the most robust determination being that of
Rutledge et al.~(1997a).  Utilizing stars from 52 Galactic globular
clusters they found a metallicity independent value of $\beta = 0.64 \pm
0.02$ \AA ~mag$^{-1}$, covering clusters in the range $-2.1 \lea$ [Fe/H]
$\lea -0.6$. Similarly, C04 found $\beta = 0.66 \pm 0.03$ for the globular
clusters in their sample.  However, when their open clusters were 
included, the slope steepened to $\beta = 0.73 \pm 0.04$.  This steepening 
of the relationship between $W'$ and $V - V_{HB}$ with [Fe/H] is in 
qualitative agreement with the theoretical results of J\o rgensen et 
al.~(1992).  
Since our target clusters span an age and metallicity range similar to the
entire calibration cluster sample observed by C04, for $\beta$ we have 
chosen to adopt their value of 0.73, which is based on both their open and 
globular calibration clusters.
To validate this approach, as 
mentioned in \S \ref{sec:acquisition}, during our science observations we 
observed a subsample of the calibration clusters used by C04 and found 
that, to within the errors, our measurements are identical to theirs, as 
is expected, given that essentially the same instrument setup was used in 
both programs.

Before proceeding to the last step of the CaT calibration, we need to
address the issue of possible age effects on these calculations.  As noted
by previous authors (e.g.~Da Costa \& Hatzidimitriou 1998; C04; Koch et
al.~2006), the age of a stellar population affects the luminosity of core
helium burning stars and may introduce systematic errors in determining
$V-V_{HB}$ and, therefore, metallicities derived via the CaT method.  
Experiments by C04 and Koch et al.~2006 have shown that age effects
brought about by using an inappropriate $V_{HB}$ for any given RGB star
will typically cause errors in [Fe/H] on the order of $\pm$ 0.05 dex, but
can, in extreme cases, be as large as $\pm$ 0.1 dex.  One can avoid this
type of uncertainty by observing populous clusters, since this allows the
correlation of a given RGB star to a specific HB/RC, which is composed of
stars of the appropriate age and, therefore, has a well defined mean
magnitude.  However, Da Costa \& Hatzidimitriou (1998) still had to
address the issue of age effects for their sample of SMC clusters due to
the fact that many of their target clusters were considerably younger than
the Galactic globular clusters used in the CaT calibration of Da Costa \&
Armandroff (1995); thus, they sought to correct for the difference in age
between the target and calibration clusters.  Using adopted cluster ages,
along with theoretical isochrones, Da Costa \& Hatzidimitriou (1998)
estimated the change in $V_{HB}$ from the old to the young populations,
thereby creating age-corrected metallicities for their targets.  Their
corrections were of the order of 0.05 dex, which is smaller than the
precision of the abundances.  In contrast to Da Costa \& Hatzidimitriou
(1998), we have made no attempt to calculate any age corrections for the
following reason.  We utilize the CaT calibration of C04, which is based
on a sample of both globular and open clusters, covering a wide range of
ages and metallicities.  With the inclusion of younger clusters, the
variation of $V_{HB}$ with age is built into the CaT calibration,
specifically in Eq.~\ref{eq:reduced} and the steeper value for $\beta$
than what has been found by authors only considering globular clusters.  
Thus, age corrections are not required for our abundance data.


Finally, Rutledge et al.~(1997b) showed that for Milky Way globular
clusters there is a linear relationship between a cluster's reduced
equivalent width and its metallicity on the Carretta \& Gratton (1997)  
abundance scale.  C04 extended this calibration to cover a larger range of
ages (2.5 Gyr $\lea$ age $\lea$ 13 Gyr) and metallicities ($-$2 $\lea
[Fe/H] \lea$ $-$0.2) than previous authors and because their calibration
is closer in parameter space to our cluster sample, we adopt their
relationship where
\begin{equation}
[Fe/H] = (-2.966 \pm 0.032) + (0.362 \pm 0.014)W'.
\label{eq:feh}
\end{equation}
We note that, while this calibration actually combines two metallicity
scales (Carretta \& Gratton 1997 for the globular clusters and Friel et
al.~2002 for the open clusters), C04 find no evidence of age effects on
the calibration or any significant deviation from a linear fit to suggest
that these two populations are not ultimately on the same [Fe/H] scale
(see their Figure 4).  Although some of our clusters are likely younger 
than the 2.5~Gyr age limit established in the calibration of C04, the CaT 
line strengths for red giants of $\sim$1~Gyr are not expected to deviate 
strongly from a simple extrapolation of the fitting formula (based on the 
empirical fitting functions from Cenarro et al.\ 2002 applied to 
isochrones published in Girardi et al.\ 2000), so we use the above 
calibration for all of our clusters.

\section{Analysis}

As was mentioned in the previous section, knowledge of the relative
brightness of each target star and the cluster HB is imperative to the
accurate calculation of $W'$ and thus [Fe/H] for each star.  To determine
$V - V_{HB}$ we utilized the preimages necessary for creating the slit
masks used by FORS2.  Small aperture photometry was performed on these $V$
and $I$-band images so as to allow us to create cluster CMDs down below
the core helium burning red clump stars.  For the younger clusters in our
sample, $V_{HB}$ was measured as the median magnitude of cluster RC stars.  
Cluster stars were isolated from the field by selecting stars within the
inner half of the apparent cluster radius.  We then placed a standard
sized box (0.8 mag in $V$ and 0.2 mag in $V-I$) around each cluster RC and
used only the stars within this box in our calculation of $V_{HB}$.  
Regarding clusters with bona fide HBs, i.~e.~old clusters, we compared our
instrumental photometry to published photometry and calculated a rough
zeropoint for our data, allowing the conversion of published $V_{HB}$
values onto our instrumental system.  Literature sources for the five old
clusters are as follows:  NGC 1841 $-$ Alcaino et al.~(1996);  NGC 2019
$-$ Olsen et al.~(1998); NGC 2257 and Hodge 11 $-$ Johnson et al.~(1999);
Reticulum $-$ Walker (1992).  Errors in $V_{HB}$ are taken as the standard
error of the median for clusters where we measured the RC directly; for
the HB in old clusters we adopt 0.1 mag.  We note that although we have
not calibrated our photometry onto a standard system, the $V-I$ color term
for the FORS2 filter system is expected to be small ($<$0.02 mag), thus
having little effect on the relative brightnesses of our target stars over
the small range of colors covered by the RGB.

\subsection{Cluster Membership}

\begin{figure}
\begin{center}
\plotone{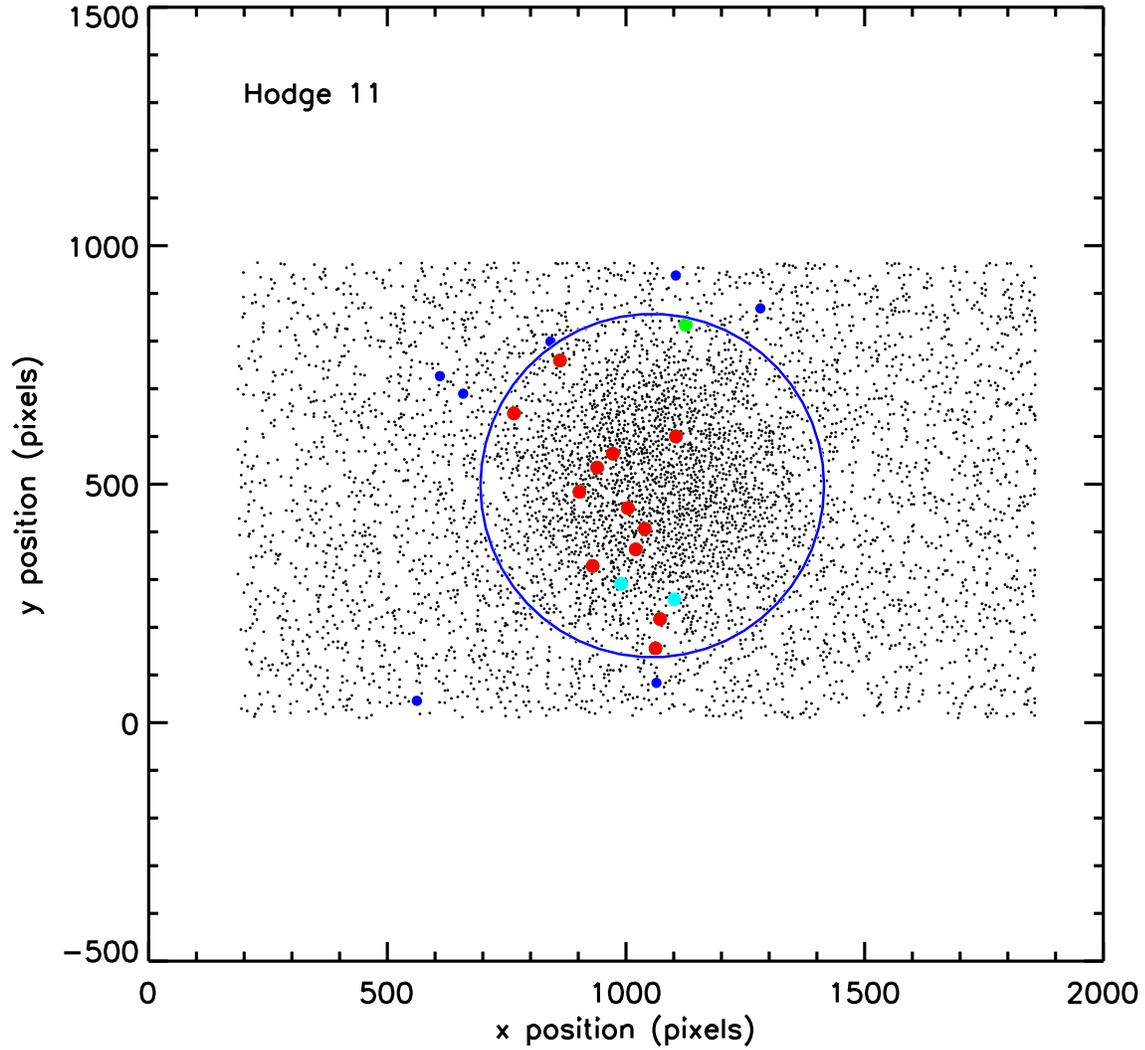}
\caption{
Shown are the $xy$ positions of our target stars (large filled points) on
the Hodge 11 field.  The adopted cluster radius is marked by the large
open circle, and stars outside of this radius are considered non-members.  
The color coding of symbols in Figs.~\ref{fig:h11_xy}$-$\ref{fig:h11_cmd}
is as follows: blue points represent non-members that are outside the
cluster radius.  Teal and green points are non-members that were cut
because of discrepant radial velocities and metallicities, respectively.  
Finally, red symbols denote cluster members.
} 
\label{fig:h11_xy}
\end{center}
\end{figure}

\begin{figure}
\begin{center}
\plotone{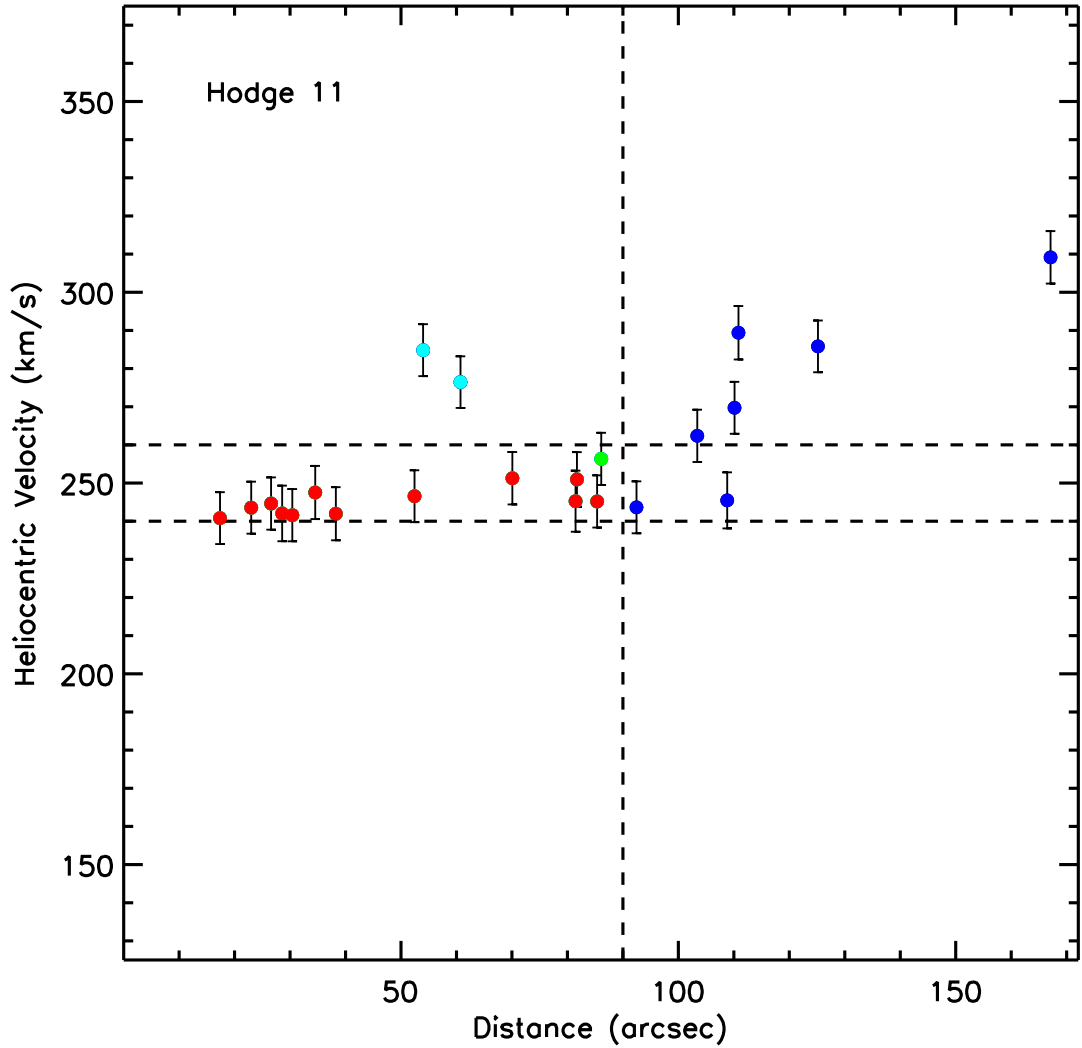}
\caption{
Radial velocities for our spectroscopic targets as a function of distance
from the Hodge 11 cluster center.  The horizontal lines represent our 
velocity cut and have a width of $\pm$10 km s$^{-1}$.  The cluster radius 
is shown as the vertical line and the color coding of points is discussed 
in Fig.~\ref{fig:h11_xy}.  The errors bars shown represent the random 
error in determining the radial velocity for each star, where we have 
added in quadrature the slit centering and cross-correlation errors.
}
\label{fig:h11_rv}
\end{center}
\end{figure}

\begin{figure}
\begin{center}
\plotone{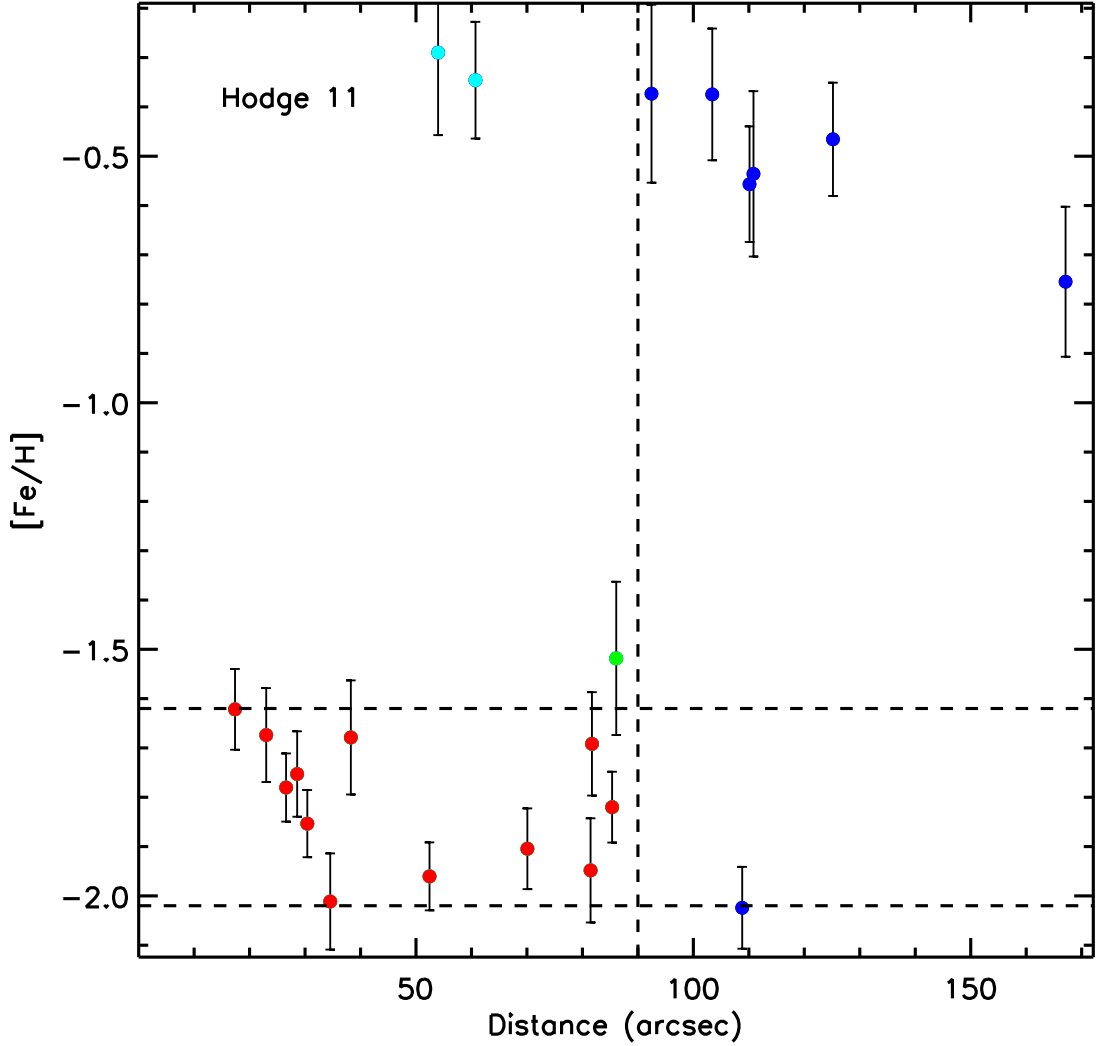}
\caption{
Metallicities are plotted as a function of position for all target stars 
in Hodge 11.  The [Fe/H] cut of $\pm$0.20 dex is denoted by the horizontal 
lines.  For this old, metal-poor cluster, the field ([Fe/H] $\sim$ $-0.5$) 
is easily distinguished from the cluster (red points).  We note that the 
color coding is the same as for previous figures.  The plotted error bars 
represent the random error in calculating [Fe/H], where we have propogated 
the error in measuring the equivalent widths through our calculations.
}
\label{fig:h11_feh}
\end{center}
\end{figure}

\begin{figure}
\begin{center}
\plotone{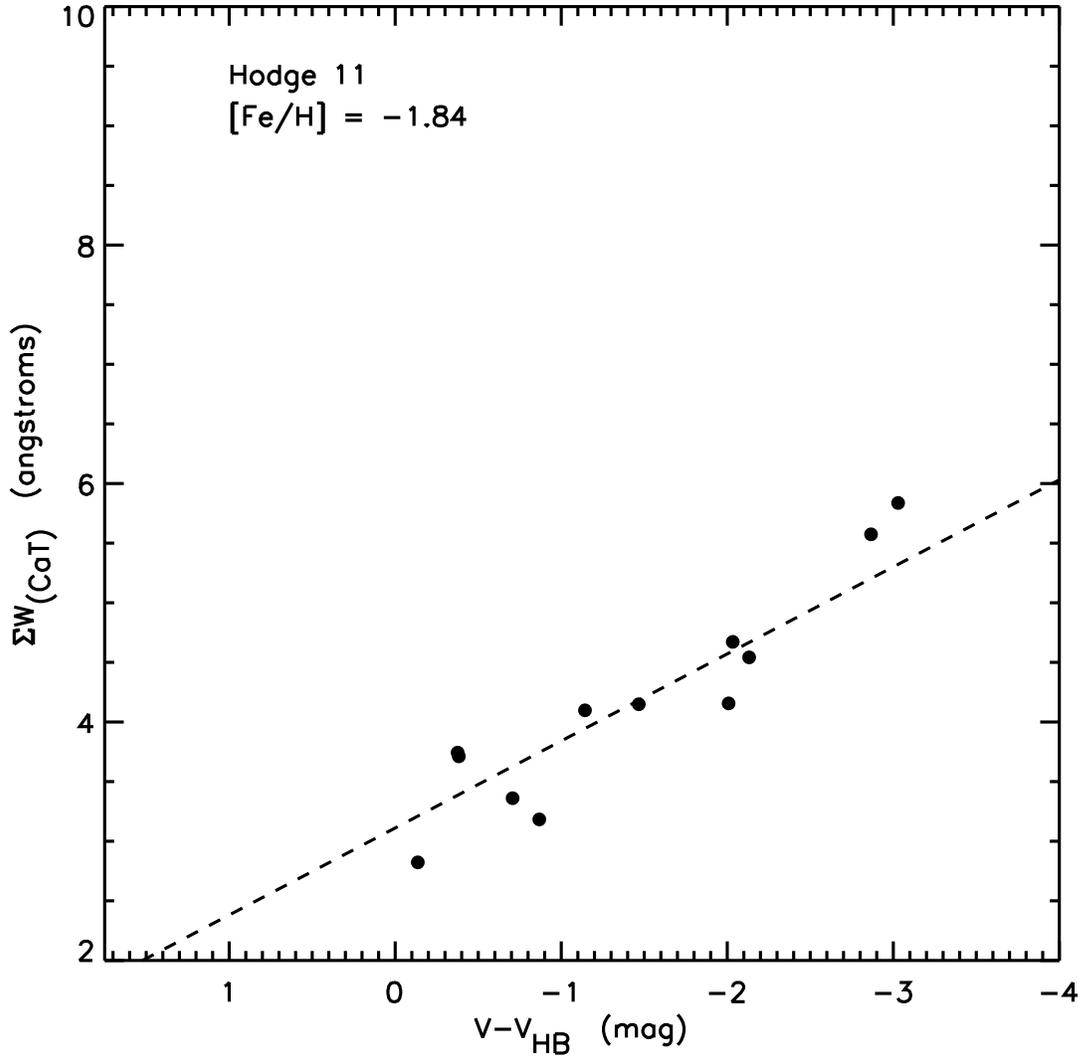}
\caption{
Summed equivalent width versus brightness above the horizontal branch is 
presented for all stars considered to be members of Hodge 11.  The dashed 
line is an isoabundance line at the mean metallicity of the cluster, 
[Fe/H] = $-$1.84, and has a slope $\beta$ = 0.73.
}
\label{fig:h11_ew}
\end{center}
\end{figure}

\begin{figure}
\begin{center}
\plotone{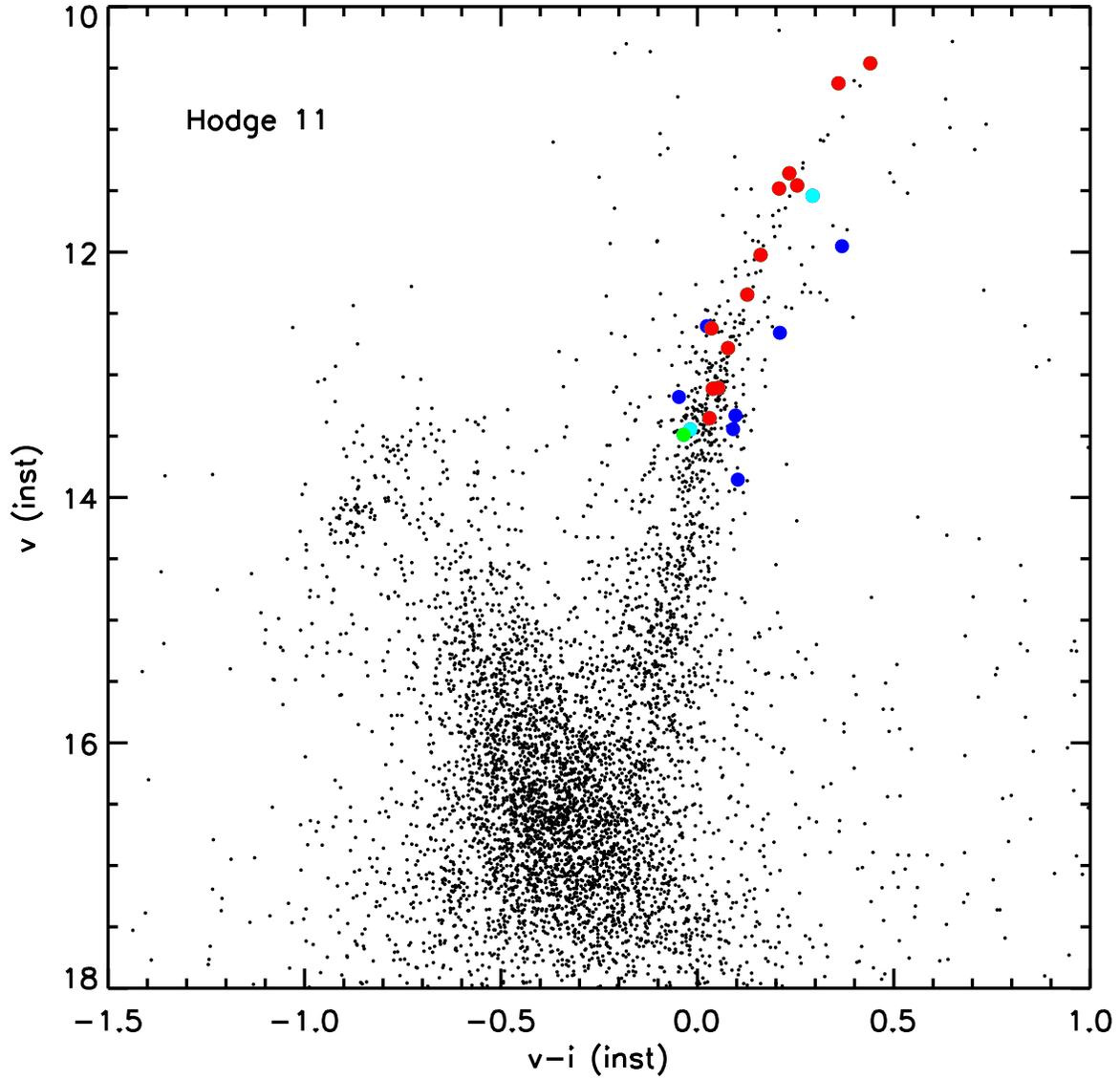}
\caption{
This figure shows the CMD for the entire Hodge 11 field, with target stars 
marked as described in Fig.~\ref{fig:h11_xy}; cluster members lie along 
the RGB and AGB.
}
\label{fig:h11_cmd}
\end{center}
\end{figure}

We use a combination of three criteria to isolate cluster members from
field stars.  This process is identical for all clusters, so we will
illustrate the process using Hodge 11.  First, the cluster centers and
radii are chosen by eye, based primarily on the photometric catalog.  As
an example, Fig.~\ref{fig:h11_xy} shows $xy$ positions for all stars
photometered in the Hodge 11 field with large filled points denoting our
target stars and the large circle representing the adopted cluster radius;
target stars marked in blue (see figure caption for a discussion of the
color coding used in Figs.~\ref{fig:h11_xy}$-$\ref{fig:h11_cmd}) are
considered non-members due to their distance from the cluster center.  We
note that stars outside of the cluster radius were observed so as to
define parameters for the LMC field, which aids in isolating cluster
members. Next, radial velocity versus position is plotted in
Fig.~\ref{fig:h11_rv}.  Stars moving at the velocity of Hodge 11 are
easily identified due to their smaller velocity dispersion and lower mean
velocity than that of the field stars.  Our velocity cut, denoted by the
horizontal lines, has been chosen to represent the expected observed
velocity dispersion in each cluster.  To determine this, we have adopted
an intrinsic cluster velocity dispersion of 5 km s$^{-1}$ and added this
in quadrature with our adopted radial velocity error, 7.5 km s$^{-1}$,
which results in an expected dispersion of $\sim$ 9 km s$^{-1}$.  Thus, we
have rounded this up and adopted a width of $\pm$ 10 km s$^{-1}$ for our
radial velocity cut. The cluster radius (vertical line) is marked for
reference.  Finally, Fig.~\ref{fig:h11_feh} shows metallicity as a
function of position for stars in Hodge 11, with horizontal lines
representing the [Fe/H] cut that has been applied to these data.  For
stars in six of our clusters we have processed both sets of spectra and
compared the two [Fe/H] measurements so as to directly determine the
metallicity error for each star.  Based on these data we find
$\sigma_{[Fe/H]} \approx$ 0.15 dex, which we adopt as the random error in
[Fe/H] for each star.  We have rounded this up to $\pm$ 0.20 dex for use
as the metallicity cut shown in Fig.~\ref{fig:h11_feh}.  Red filled points
denote stars that have made all three cuts and are therefore considered to
be cluster members.  Since we had no a priori membership information, up
to this point we have used a value for $V_{HB}$ which was derived from the
entire field, rather than just the cluster.  Thus, we have recalculated
$W'$ (and [Fe/H]) using the appropriate cluster $V_{HB}$ value.  In
Fig.~\ref{fig:h11_ew} we present the traditional $\Sigma W$ versus $V -
V_{HB}$ plot for cluster members with the dashed line representing the
mean metallicity of Hodge 11.  The CMD in Fig.~\ref{fig:h11_cmd} shows all
stars photometered in the Hodge 11 field; cluster members (red points) lie
on the RGB and AGB.  

In Table \ref{table:all_stars}, for all stars determined to be members of
the observed LMC clusters, we list the following information:  stellar
identification number, right ascention and declination (as determined from
the preimages), heliocentric radial velocity and its associated error,
$V-V_{HB}$, and $\Sigma W$ along with the error in measuring $\Sigma W$.

\subsection{Cluster Properties}
\label{sec:properties}

Cluster properties derived from our data are presented in Table
\ref{table:properties}, with the number of cluster stars given in column
2, the mean heliocentric radial velocities and mean metallicities in
columns 3 and 5 and their respective standard error of the mean values in
columns 4 and 6.  For the clusters SL 4, SL 41, SL 396, Hodge 3, SL 663,
and SL 869, we report the first spectroscopically derived metallicity and
radial velocity values based on individual stars within these clusters.  
Additionally, NGC 1718 and NGC 2193 have no previously reported
spectroscopic [Fe/H] values, however OSSH derived velocities for these two
clusters.  Of these eight clusters, NGC 1718 occupies a particularly
interesting area of parameter space as it is the most metal poor of our
intermediate age clusters, with a metallicity comparable to that of ESO
121 (see discussion in appendix). As mentioned previously, we have not
derived values for NGC 1861 since it appears to be younger than 1 Gyr (see
appendix).

\subsubsection{Metallicities}

\begin{figure}
\begin{center}
\plotone{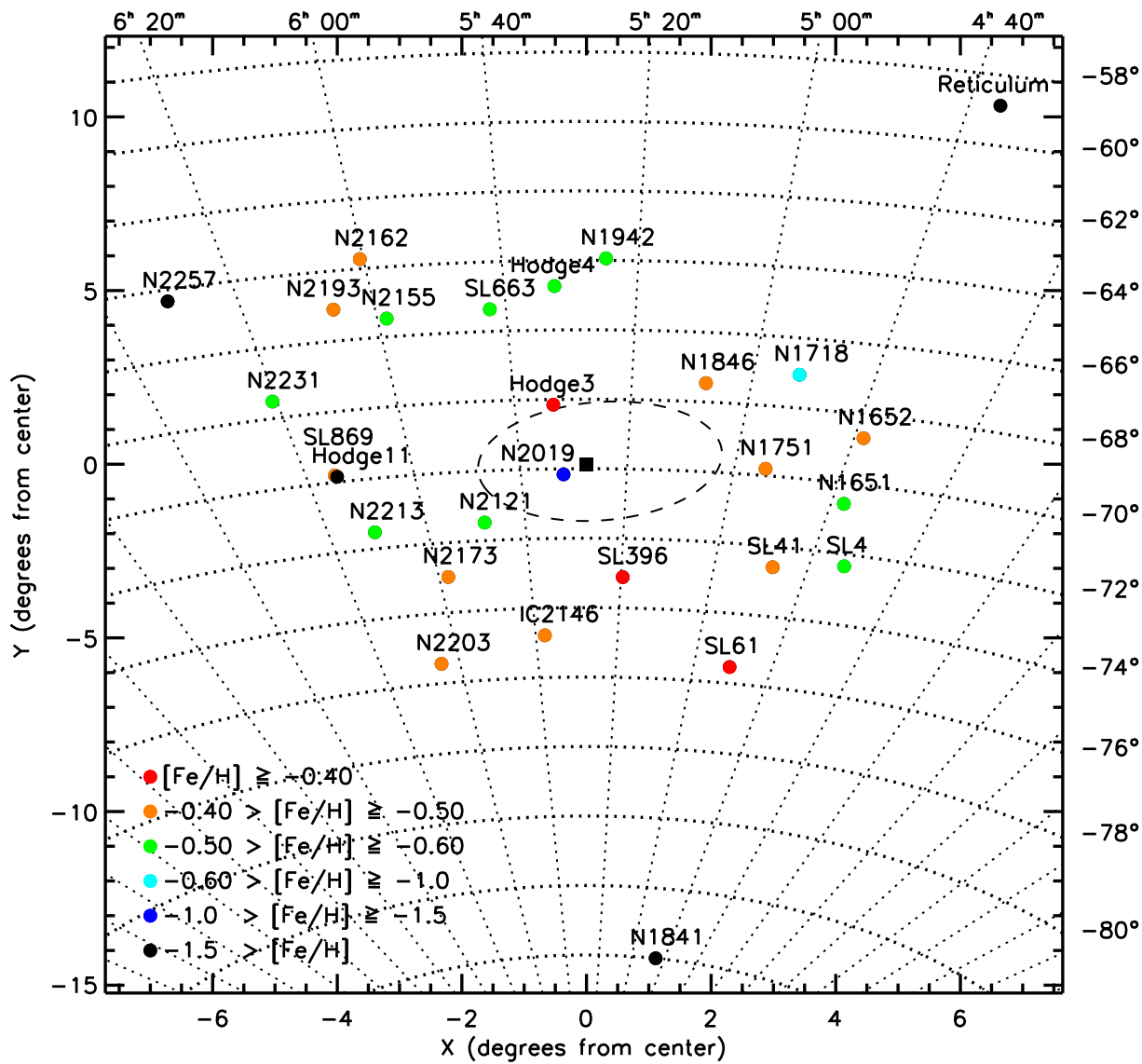}
\caption{
Shown are the positions on the sky and derived metallicities for our 
target clusters; metallicity bins are given in the lower left corner 
of the plot.  The adopted LMC center is marked with the filled square and 
the dashed line roughly outlines the bar.  See \S 3.2 for a detailed 
discussion.
}
\label{fig:cluster_pos}
\end{center}
\end{figure}

\begin{figure}
\begin{center}
\plotone{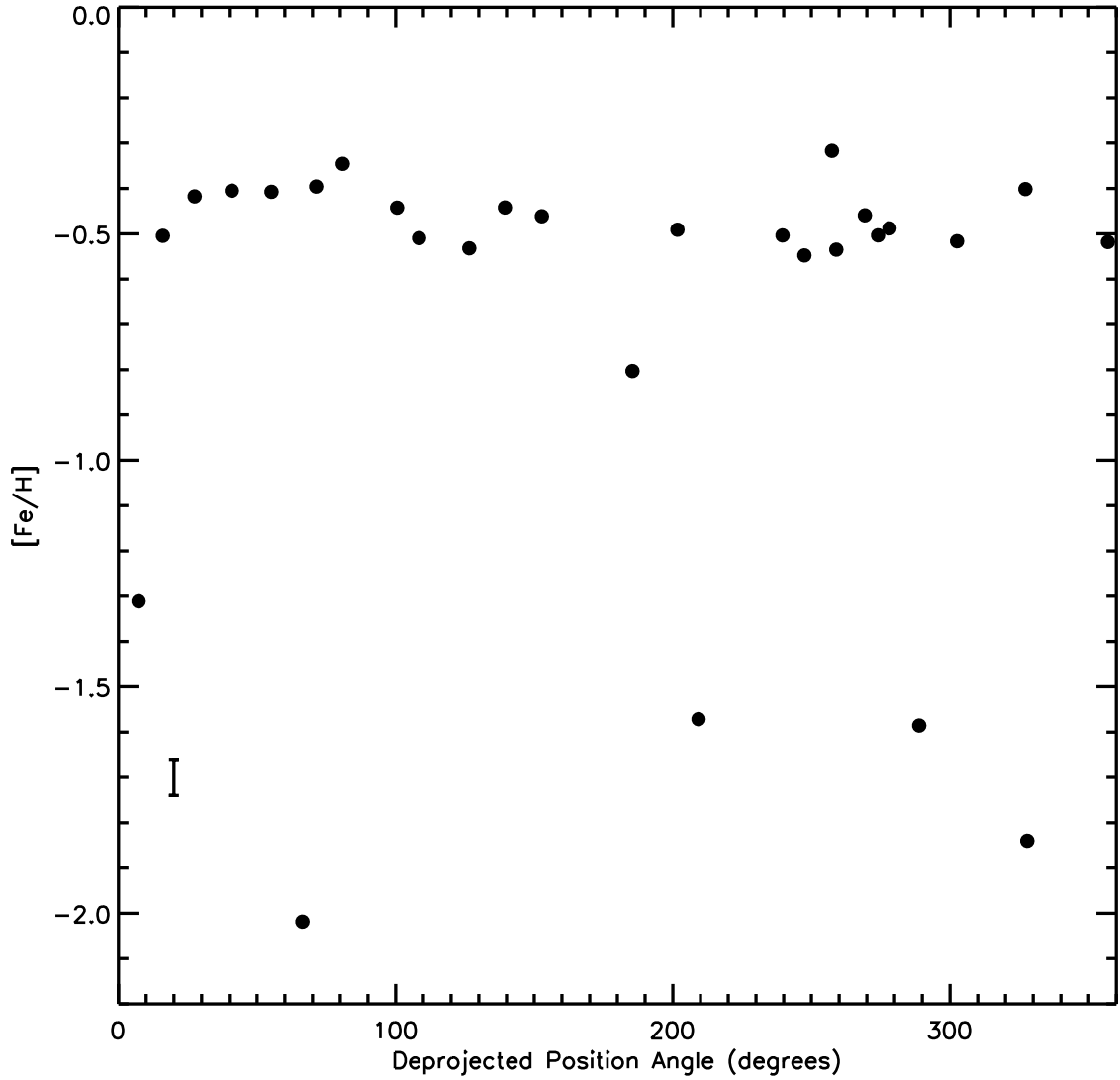}
\caption{
We plot the metallicities for our target clusters as a function of 
deprojected position angle, where we have used the LMC geometry of van der 
Marel \& Cioni (2001) to correct for projection effects.  This plot 
illustrates that there is no apparent relation between position angle 
and metallicity in the LMC.  The error bar shown in the lower lefthand 
corner of the plot illustrates the average random error in [Fe/H].
}
\label{fig:feh_vs_pa}
\end{center}
\end{figure}

\begin{figure}
\begin{center}
\plotone{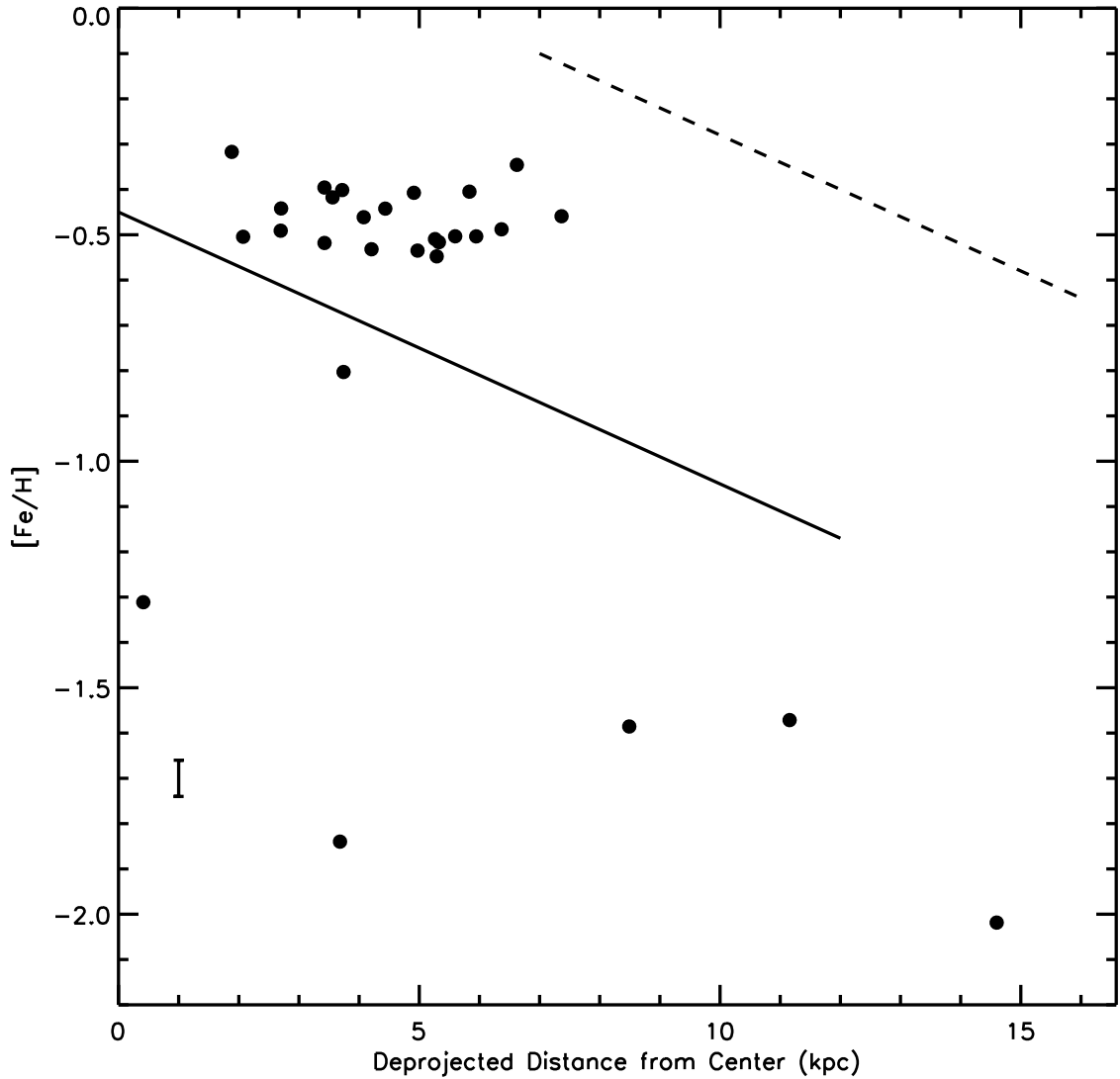}
\caption{  
Cluster metallicities are plotted as a function of deprojected distance
(in kpc) from the center of the LMC; we have assumed a distance of
$(m-M)_0 = 18.5$.  Overplotted are the metallicity gradients observed in
the MW open clusters (dashed line; Friel et al.~2002) and M33 (solid line; 
Tiede et al.~2004), which help to further illustrate that the LMC's 
cluster system lacks the metallicity gradient typically seen in spiral
galaxies.  This flattened gradient is likely caused by the presence of the
central bar (Zaritsky et al.~1994).  As with the previous plot, the 
average random error is illustrated by the error bar on the lower left.  
}
\label{fig:feh_vs_dist}
\end{center}
\end{figure}

Positions on the sky for each cluster are shown in
Fig.~\ref{fig:cluster_pos}, along with the metallicity bin into which each
cluster falls, represented by the color of the plotting symbol.  For two
of the higher metallicity bins (orange and green points), the bin size is
roughly twice the standard error in [Fe/H], so it is possible that cluster
errors could `move' clusters between these and adjacent bins. The adopted
center of the LMC ($\alpha = 5^h27^m36^s$, $\delta =
-69{\degr}52{\arcmin}12{\arcsec}$; van der Marel et al.~2002) is marked by
the filled square and the dashed oval represents the 2$\degr$
near-infrared isopleth from van der Marel (2001), which roughly outlines
the location of the LMC bar.  Conversion from RA and Dec to Cartesian
coordinates was performed using a zenithal equidistant projection
(e.g.~van der Marel \& Cioni 2001, their equations 1$-$4); for reference,
lines of RA and Dec are marked with dotted lines.  In
Figs.~\ref{fig:feh_vs_pa} and \ref{fig:feh_vs_dist}, we further explore
the metallicity/position relationship for LMC clusters by plotting
metallicity as a function of deprojected position angle and radial
distance (in kpc), respectively. We have corrected for projection effects
by adopting 34$\fdg$7 as the inclination and 122$\fdg$5 for the position
angle of the line of nodes of the LMC (van der Marel \& Cioni 2001).  In
this rotated coordinate system, a cluster with a position angle of zero
lies along the line of nodes, and angles increase counterclockwise; for
reference, NGC 2019 has a position angle of $\sim$ 8$\degr$. Radial
distances were converted from angular separation to kiloparsecs by
assuming an LMC distance of $(m-M)_0 = 18.5$ ($\sim$ 50 kpc); at this
distance, one degree is $\sim$ 870 pc.  Combined, these three figures
illustrate that, similar to what was found by OSSH (and also Geisler et
al.~2003), there is no [Fe/H] gradient in terms of either position angle
or radial distance for the higher metallicity clusters in our sample.  
While we cannot make strong comments on the metal poor clusters due to our
small sample size, it is well known that a number of metal poor clusters
([Fe/H] $\lea -1.5$) exist in the inner portions of the LMC (e.g.~OSSH),
suggesting that neither the Population I nor the Population II clusters
exhibit a metallicity gradient.

In Fig.~\ref{fig:feh_vs_dist} we have over plotted both the MW open
cluster metallicity gradient from Friel et al.~(2002; dashed line) and the 
M33 gradient from Tiede et al.~(2004; solid line).  Neither of these disk
abundance gradients resembles what we see among the LMC clusters. The
question of how to interpret this difference takes us to the work of
Zaritsky, Kennicutt, \& Huchra (1994). They studied the \ion{H}{2} region
oxygen abundances in 39 disk galaxies. Their data suggest that disk
abundance gradients are ubiquitous in spiral galaxies. However, the
presence of a classical bar in the galaxy - one that extends a significant
fraction of the disk length - tends to weaken the gradient. This
observation seems to find support in the appearance of
Fig.~\ref{fig:feh_vs_dist}. In the case of the LMC, the presence of a
strong bar component may have diluted the metallicity gradient originally
present in the star clusters leading to a cluster population that is well
mixed.
We note that this result is also consistent with the conclusion of 
Pagel et al.~(1978), who found little evidence for a gradient in oxygen 
abundance based on a survey of \ion{H}{2} regions within 4 kpc of the LMC 
center.  The Pagel result, that $d \log$(O/H)/$d$R $= -0.03 \pm 0.02$ 
dex kpc$^{-1}$, parallels our non-detection of a gradient in cluster 
metallicities.

\subsubsection{Kinematics}

To characterize the rotation of their clusters, Schommer et al.~(1992) fit 
an equation of the form
\begin{equation} 
V(\theta) = \pm V_m\{[\tan(\theta - \theta_{0}\sec i^2] + 1\}^{-0.5} + 
V_{sys} 
\label{eq:rotation}
\end{equation} 
to their radial velocity data using a least-squares technique to derive
the systemic velocity ($V_{sys}$), amplitude of the rotation velocity
($V_m$), and the orientation of the line of nodes ($\theta_0$);  they
adopted an inclination of $27\degr$.  Their best fit parameters give a
rotation amplitude and dispersion consistent with the LMC clusters having
disk-like kinematics, with no indications of the existence of a pressure
supported halo. We note that, due to the non-circularity of the LMC,
$\theta_0$ in Eq.~\ref{eq:rotation} is not the true orientation of the
line of nodes (the intersection of the plane of the sky and the plane of
the LMC), but rather it marks the line of maximum velocity gradient (van
der Marel \& Cioni 2001).  More recently, van der Marel et al.~(2002) used
velocities of 1041 carbon stars to study kinematics in the LMC. Similarly,
they found that these stars exhibit a disk-like rotation with $V/\sigma$ =
2.9 $\pm$ 0.9, suggesting that these stars reside in a disk that is
slightly thicker than the Milky Way thick disk ($V/\sigma$ $\approx$ 3.9).

\begin{figure}
\begin{center}
\plotone{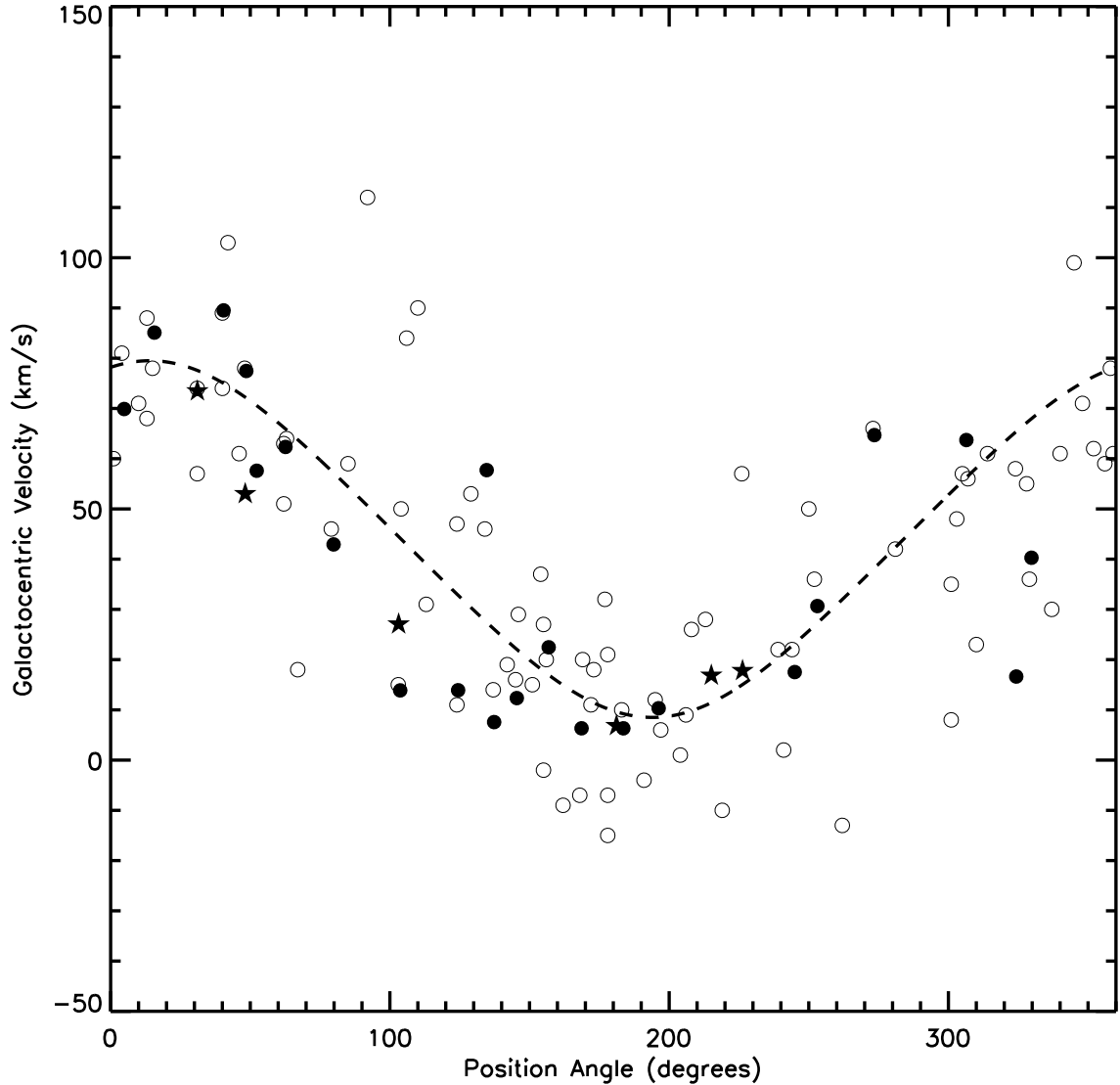}
\caption{
Galactocentric radial velocities as a function of position angle on the
sky are plotted for the clusters in our sample (filled symbols) as well as
those from Schommer et al.~(1992; open circles);  the six clusters in our
sample with no previous velocity determinations are plotted as filled
stars and all others in our sample are filled circles.  Rotation curve
solution number 3 from Schommer et al.~(1992) is overplotted as the dashed
line, showing that both data sets are consistent with circular rotation.  
We note that we have not plotted a representative error bar since our 
plotting symbols are roughly the same size as the average random velocity 
error.
}
\label{fig:cluster_rot}
\end{center}
\end{figure}

In Fig.~\ref{fig:cluster_rot}, we have plotted Galactocentric radial
velocity versus position angle on the sky for our sample, along with
velocity data for all clusters listed in Schommer et al.~(1992).  So as to
be consistent with the approach of Schommer et al.~(1992), we adopt the
Galactocentric velocity corrections given by Feitzinger \& Weiss (1979).  
Additionally, for this figure only, we have adopted their LMC center
($\alpha = 5^h20^m40^s$, $\delta = -69{\degr}14{\arcmin}10{\arcsec}$;
J2000) for use in calculating the position angles of our clusters.  We
have used the standard astronomical convention where North has a position
angle of zero and angles increase to the East; NGC 1942 has a position
angle of $\sim 4\degr$ in this coordinate system.  Data from Schommer et
al.(1992) are plotted as open circles and our data are plotted as filled
stars for the clusters with previously unpublished velocities and filled
circles for the remainder of our clusters; overplotted on this figure
(dashed line) is the rotation curve solution number 3 from Schommer et
al.~(1992).  For the clusters in common between these two data sets, we
find excellent agreement, with a mean offset of 0.15 km s$^{-1}$, where
our velocities are faster than those of Schommer et al.~(1992).  
Additionally, the derived velocities for the six `new' clusters show that
their motions are consistent with the findings of Schommer et al.~(1992)
in that the LMC cluster system exhibits disk-like kinematics that are very
similar to the HI disk, and has no obvious signature of a stellar halo.



\section{Comparison with Previous Work}
\label{sect:comparison}

As mentioned in the introduction, OSSH and Suntzeff et al.~(1992) have
provided the only previous large scale, spectroscopic [Fe/H] calculations
for clusters in the LMC.  Similar to our work, they utilized the CaT lines
as a proxy for measuring Fe abundance directly, but with two important
differences: they used the absolute magnitude of their stars, based on the
spectral intensity at 8600 \AA , as a surface gravity estimator instead 
of $V-V_{HB}$, and their [Fe/H] calibration is based largely on the Zinn 
\& West (1984) metallicity system, with the addition of two open clusters
that have metallicities derived from various spectrophotometric indices
(see their table 7).  This will introduce two systematic offsets that make
it inappropriate to directly compare the OSSH values to our work and other
recently measured CaT abundances: first, the use of M$_{8600}$ creates a
dependence on the relative distances of the calibrating clusters and the
LMC, and the globular cluster distance scale has been much-revised in the
post-Hipparcos era (Reid 1999); and second, it has been shown (e.g.,
Rutledge et al.~1997b) that the Zinn \& West scale is non-linear compared
to the more recent Carretta \& Gratton (1997) scale based on
high-resolution spectra of globular cluster red giants.  To put the work
of OSSH on the Carretta \& Gratton system, Cole et al.~(2005) perform a
non-linear least-squares fit to calibration clusters in common between
their work and that of OSSH.  They find that one can estimate the
abundance of OSSH clusters on the metallicity system we have used via the
following conversion:
\begin{equation}
[Fe/H] \approx -0.212 + 0.498[Fe/H]_{OSSH} - 0.128[Fe/H]^{2}_{OSSH}.
\label{eq:ossh_conversion}
\end{equation}
This equation approximates the metallicity that OSSH {\it would} have
derived from their spectroscopic data and calibration procedure, but with
updated metallicities for their calibration clusters; it does not attempt
to account for any other differences in the treatment of the data.

\begin{figure}
\begin{center}
\plotone{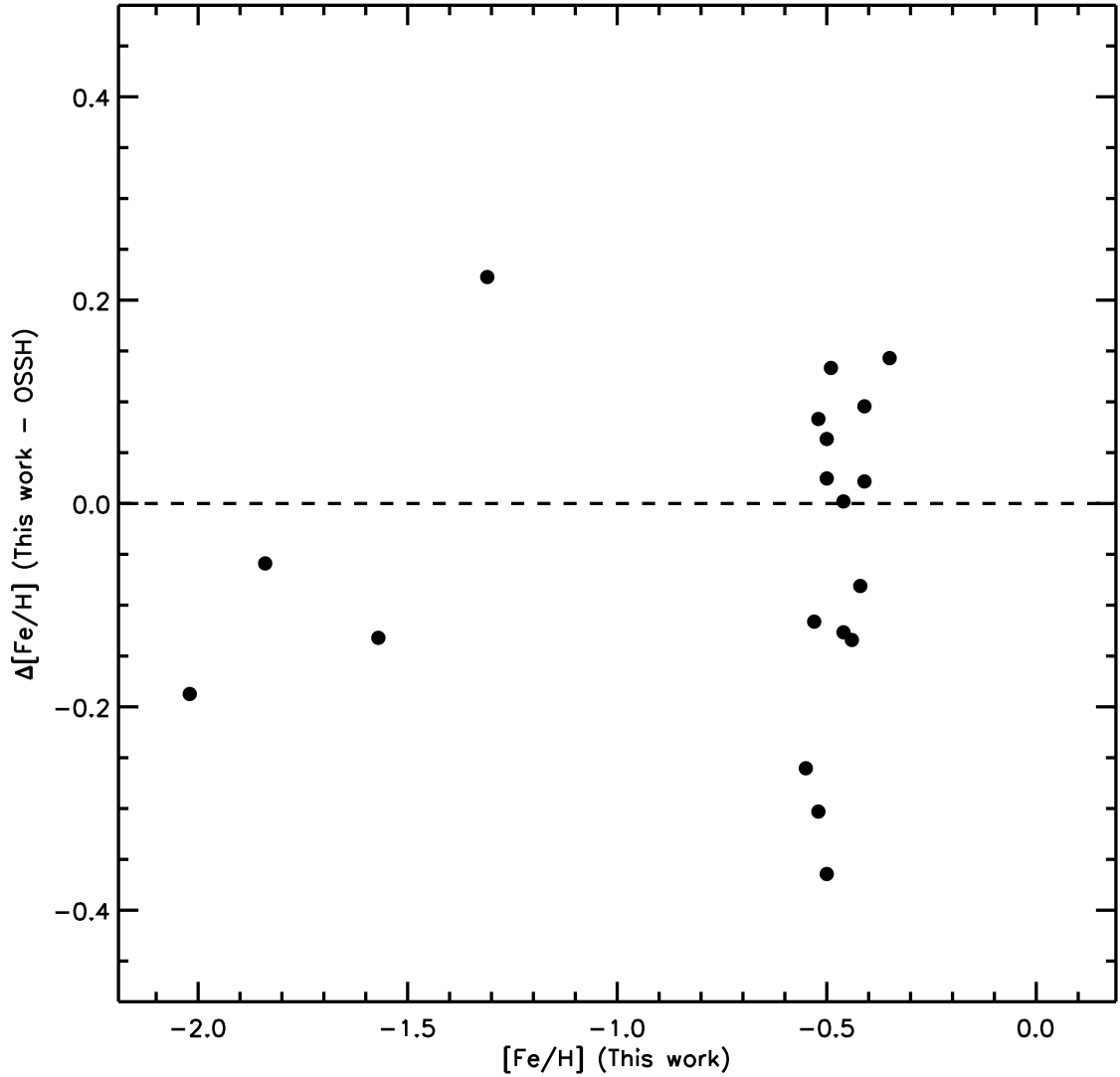}
\caption{Derived metallicities for clusters in common between our study
and that of OSSH are compared; we note that [Fe/H] values from OSSH were
converted onto the metallicity scale we have used via
Eq.~\ref{eq:ossh_conversion}.  This comparison shows that, to within the
errors, there is relatively good agreement between our results and those
of OSSH (see \S \ref{sect:comparison} for more details).
}
\label{fig:compare_ossh}
\end{center}  
\end{figure}

In columns 3 and 4 of Table \ref{table:other_data} we list [Fe/H] for
clusters in OSSH and Suntzeff et al.~(1992) in common with our target
clusters, where column 3 gives their published values and in column 4 we
have converted their numbers onto our metallicity system using
Eq.~\ref{eq:ossh_conversion};  the number of stars used by OSSH in
calculating final cluster metallicities is given in parenthesis in column
4 and our derived metallicities are given in column 2 for reference. In
Fig.~\ref{fig:compare_ossh} we plot the difference between our
metallicities and their converted [Fe/H] values as a function of our
metallicities.  OSSH give their [Fe/H] errors for an individual star as
0.2 dex, therefore deviations between these data sets as large as $\pm$
0.2 are not unexpected, suggesting that these results are in relative
agreement, with no offset.  We note, however, that even with the use of
Eq.~\ref{eq:ossh_conversion}, it is very difficult to directly compare the
derived cluster abundances because of the differences in target selection
and calibration strategy.

\begin{figure}
\begin{center}
\plotone{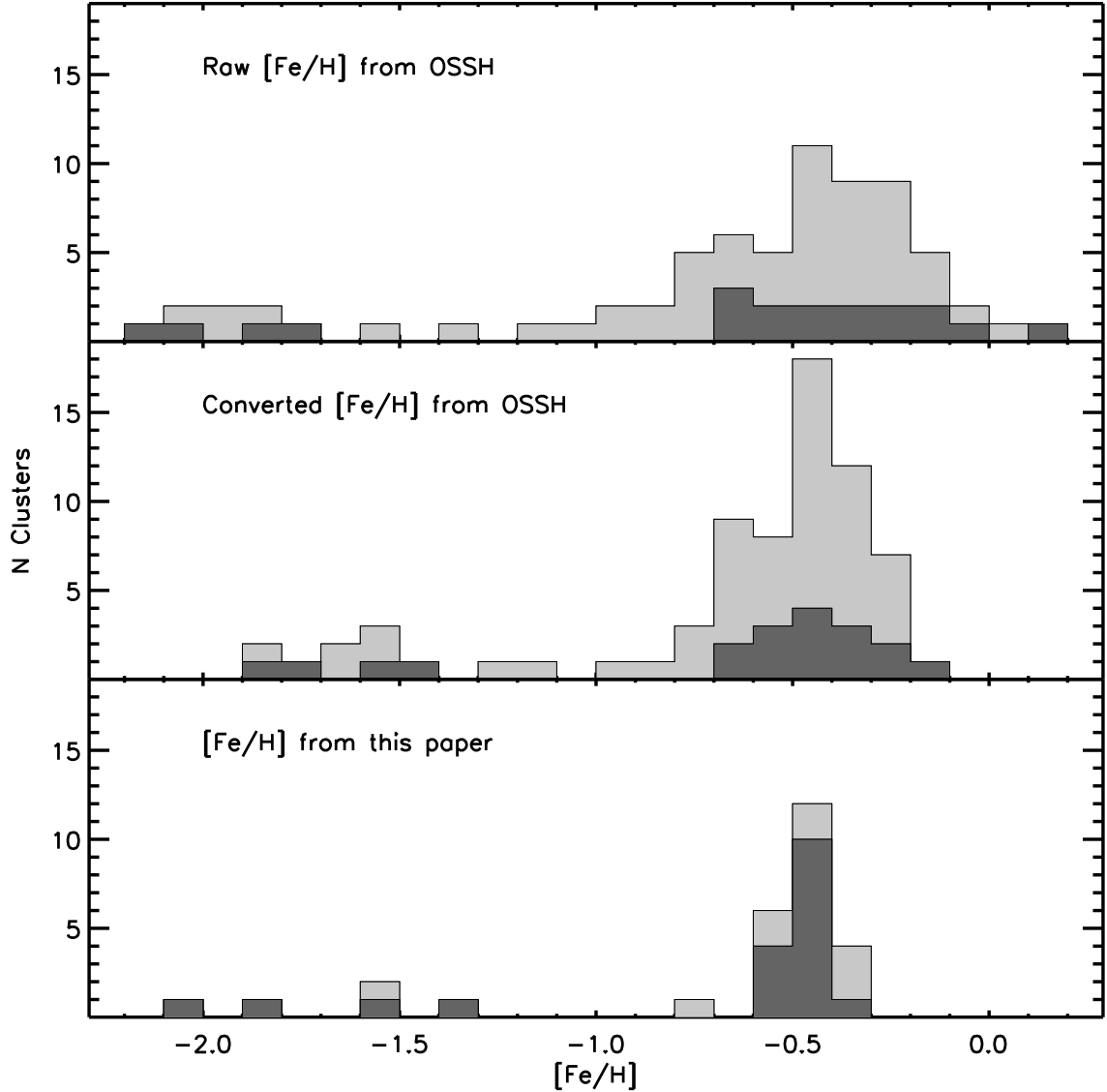} 
\caption{We plot the metallicity distribution of LMC clusters as 
determined by OSSH and this paper.  Published values from OSSH are given 
in the top panel while the middle panel shows their values converted onto 
our metallicity scale using Eq.~\ref{eq:ossh_conversion}; in the bottom 
panel we have plotted our results.  In all three panels, the dark shaded 
region shows the distribution for the 20 clusters in common between 
OSSH and this paper, while the light shaded region shows the entire 
cluster sample from each study.  Our results indicate that the LMC's 
intermediate age cluster metallicity distribution is actually much tighter 
than suggested by the results of OSSH.
}
\label{fig:clust_histog}
\end{center}
\end{figure}

While a direct comparison of [Fe/H] values is difficult, we can readily
compare the metallicity distributions of these two data sets.  As such, in
Fig.~\ref{fig:clust_histog}, we have plotted the metallicity distribution
of OSSH's raw data (top panel), converted [Fe/H] values (middle panel),
and our results (bottom panel).  The dark shaded histogram shows only the
20 clusters in common between the three panels, while the lighter
histogram plots all clusters in each sample.  From this figure it is clear
that both the raw and converted OSSH samples show an extended distribution
of intermediate metallicity clusters, whereas our cluster sample exhibits
a very tight distribution.  For the 20 clusters in common, we find a mean
[Fe/H] = $-0.47$ with $\sigma = 0.06$, while the converted OSSH
metallicities give [Fe/H] = $-0.42 \pm 0.14$.  Our tight metallicity
distribution, with a lack of higher metallicity clusters ([Fe/H] $\gea
-0.30$), is an important feature of our data for the following reason.  
Chemical evolution models suggest that metallicity is a rough estimator of
age, in that younger stellar populations should be more metal rich than
older populations since there has been more time to process material and
enrich the ISM.  Thus, intermediate age clusters should be more metal poor
than younger stellar populations in the LMC.  However, some
intermediate-age clusters in the sample of OSSH appeared to be more
metal-rich than much younger stellar populations in the LMC, which would
indicate the presence of a large spread of metallicities at any given age.  
In Table \ref{table:metal_compare} we give the mean metallicity and spread
of our entire sample of intermediate age clusters and all clusters in OSSH
with converted metallicities above $-1.0$ dex, along with published
results for a sample of younger stellar populations (e.g.~B dwarfs,
Rolleston, Trundle, \& Dufton 2002; Cepheid variables, Luck et al.~1998;
young red giants, Smith et al.~2002) and intermediate age RGB field stars
in the LMC bar (Cole et al.~2005).  This table shows that, as we would
expect from chemical enrichment models, the intermediate age clusters are
slightly more metal poor than the younger populations in the LMC.  Thus,
the much tighter metallicity distribution seen in our clusters is in
excellent agreement with the expected chemical enrichment pattern in the
LMC and alleviates the problem created by the high metallicity tail of
intermediate age clusters in the OSSH results.  Additionally, Table
\ref{table:metal_compare} shows that our intermediate age clusters have a
mean metallicity and distribution similar to that of the metal rich
component of the bar field studied by Cole et al.~(2005).  The similarity
between these two populations is in good agreement with the models of
Bekki et al.~(2004), in which the formation of the LMC bar and the restart
of cluster formation (end of the `age-gap') are both a result of the same
very close encounter with the SMC.

Finally, in Table \ref{table:other_data} we have also included [Fe/H]
values derived from high resolution spectra for NGC 1841 and NGC 2257 from
Hill (2004) and NGC 2019 and Hodge 11 from Johnson et al.~(2006).  For the
two clusters from Johnson et al.~(2006), we list [Fe/H] values that are
the average of their metallicities determined from \ion{Fe}{1} and
\ion{Fe}{2} lines, and the number of stars observed in each cluster is
given.  Two clusters, NGC 1841 and NGC 2019, show good agreement between
our metallicities, calculated from the CaT lines, and from fitting to high
resolution spectra.  In contrast, Hodge 11 and NGC 2257 show a roughly 0.3
dex offset between these methods in the sense that our values are more
metal rich than the results from high resolution spectra.  Similarly, a
preliminary result for ESO 121, which is more metal rich than the
aforementioned clusters, suggests an offset in the same direction, where
the CaT method gives a [Fe/H] value higher than what is measured with high
resolution spectra (Cole, private communication).  It has been suggested
that variations in [Ca/Fe] between calibrating clusters in the MW and
target clusters in the LMC may cause a breakdown in the utility of CaT
lines as a metallicity indicator.  However, abundances based on high
resolution spectra show that [Ca/Fe] is typically lower for LMC cluster
giants than in MW giants of the same [Fe/H], which is in the opposite
direction of what is needed to explain the difference between CaT and high
resolution results.  We also note that, for low metallicity stars,
previous authors have shown that metallicities derived from
high-resolution spectra can vary considerably (0.3 dex is not uncommon),
depending on which ionization stages, what temperature scale, and what
model atmospheres are being used (e.g.~Johnson et al.~2006, Kraft \& Ivans
2003).


\section{Summary}

As mentioned in the introduction, determining abundances for populous
clusters within the LMC is an important step in understanding the history
of this satellite galaxy.  Accurate [Fe/H] values help to break the
age-metallicity degeneracy that arises when trying to fit theoretical
isochrones to cluster CMDs, which will allow the unequivocal determination
of cluster ages, thereby providing a clear picture of the LMC's cluster
age-metallicity relation.  These clusters also serve to fill a region of 
the age-metallicity plane that is void of MW clusters;  this makes the 
LMC cluster system an important testbed for a variety of stellar 
population models.  Additionally, in a previous paper (Grocholski
\& Sarajedini 2002), we have shown that knowledge of a cluster's age and
metallicity is essential to predicting the $K$-band luminosity of the RC
for use as a standard candle.  In a future work, we will use the
metallicities derived herein to determine distances to individual populous
LMC clusters, which will allow us to compare the cluster distribution to
the LMC geometry calculated from field stars (e.g.~van der Marel \& Cioni 
2001).

In this paper we have presented the results of our spectroscopic study of
the near infrared Calcium II Triplet lines in individual RGB stars in 28
populous LMC clusters.  Utilizing the multi-object spectrograph, FORS2, on
the VLT, we have been able to determine membership and calculate
metallicities and radial velocities for, on average, 8 stars per cluster,
with small random errors (1.6 km s$^{-1}$ in velocity and 0.04 dex in
[Fe/H]).  The number of cluster members observed, combined with the
updated CaT calibration of C04 (they extended the calibration to younger
and more metal rich clusters than previous work), has allowed us to
improve upon the work of OSSH, which is the only previous large scale
spectroscopic study of individual cluster stars within the LMC.  The main
results of our paper are as follows:


1.  We report the first spectroscopically derived metallicities and radial
velocities for the following clusters: SL 4, SL 41, SL 396, SL 663, SL 
869, and Hodge 3.  In addition, NGC 1718 and NGC 2193 have no previously 
reported spectroscopic [Fe/H] values.  

2.  NGC 1718 is the only cluster in our sample that falls into the range 
$-1.3 \le$ [Fe/H] $\le -0.6$.  This metallicity region corresponds to 
the well known 3$-$13 Gyr `age gap,' within which there is only one 
cluster, ESO 121.  However, unlike ESO 121, the CMD of NGC 1718 suggests 
an age ($\sim$ 2 Gyr) much younger than the `age-gap'; we use archival 
HST/WFPC2 photometry to investigate this point in the appendix.  This age 
makes NGC 1718 one of the most metal poor intermediate age clusters in the 
LMC.  


3.  The intermediate age clusters in our sample show a very tight
distribution, with a mean metallicity of $-0.48$ dex ($\sigma$ = 0.09) and
no clusters with metallicities approaching solar.  While this is in
contrast to previous cluster results, it suggests that the formation
history of the bar (mean [Fe/H] = $-0.37$, $\sigma = 0.15$; Cole et
al.~2005) is very similar to that of the clusters.  This agrees well with
the theoretical work of Bekki et al.~(2004), which indicates that a close
encounter between the LMC and SMC caused not only the restart of cluster
formation in the LMC, but the generation of the central bar as well.  

4.  Similar to previous work, we find no evidence for the existence of a
metallicity gradient in the LMC cluster system.  This is in stark contrast
to the stellar populations of both the MW and M33, which show that
metallicity decreases as galactrocentric distance increases; the LMC's 
stellar bar is likely responsible for the well mixed cluster system.  

5.  We find that our derived cluster velocities, including the six `new'
clusters, are in good agreement with the results of Schommer et al.~(1992)
in that the LMC cluster system exhibits disk-like rotation with no
clusters appearing to have halo kinematics.

6.  Comparing our results for four clusters to [Fe/H] values recently
derived through high-resolution spectra, we find that two of the four
clusters are in good agreement, while the other two have [Fe/H] values
derived via the CaT method that are $\sim$ 0.3 dex more metal rich than
what is found from high-resolution spectra; a similar effect is seen in
preliminary results for an additional two LMC clusters.  The source of
this difference is unclear, however, it is not immediately explained by
variations in [Ca/Fe] between the CaT calibration clusters in the MW and
the LMC target clusters.  Further high resolution studies, especially 
covering the LMC's intermediate age clusters, are needed to fully address 
this issue.  


\acknowledgments

This work is based on observations collected at the European Southern
Observatory, Chile, under proposal 074.B-0417.  Pre-imaging data were
taken in service mode, thanks to the work of the Paranal Science
Operations Staff.  We would like to thank an anonymous referee for
comments that helped to improve the clarity of this manuscript.  AJG was
supported by NSF CAREER grant AST-0094048 to AS.  AAC was supported by a
fellowship from the Netherlands Research School for Astronomy (NOVA).  DG
gratefully acknowledges support from the Chilean {\sl Centro de Astrof\`\i
sica} FONDAP No.~15010003.  VVS has been supported by the NSF through
grant AST03-07534.



{\it Facilities:} \facility{VLT:Antu (FORS2)}.

\appendix

\section{Notes on Individual Clusters}

\subsection{NGC 1718}

\begin{figure}
\begin{center}
\plotone{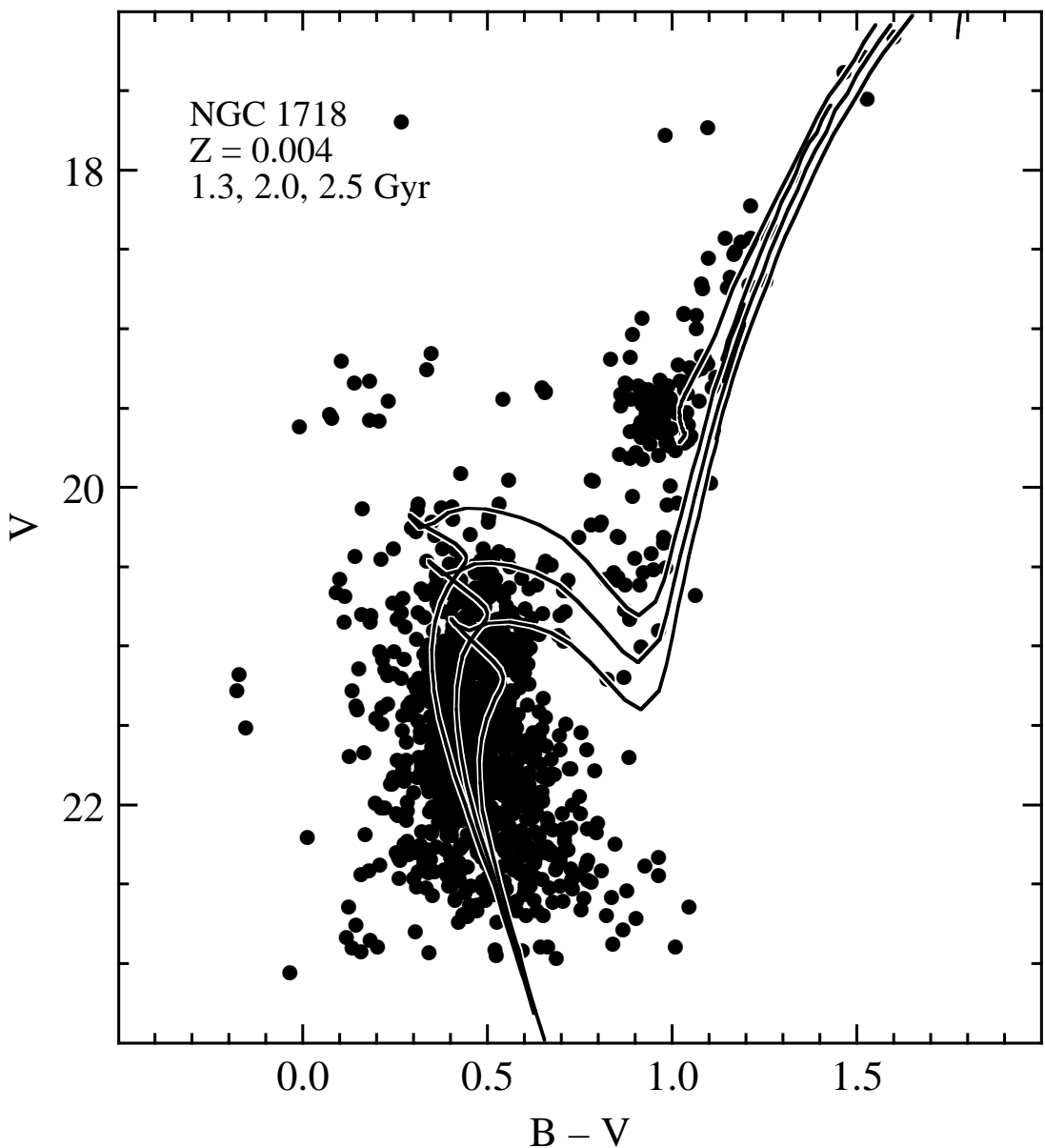}
\caption{  
Presented is the cluster CMD for NGC 1718, based on aperture photometry of
archival HST/WFPC2 images.  We overplot isochrones of 1.3, 2.0, and 2.5
Gyr (top to bottom) from Girardi et al.~(2002) that have a metallicity
($\sim$ $-$0.7 dex) similar to the value we have derived for this cluster
($-$0.8 dex).  Although this cluster has a metallicity similar to that of
ESO 121, the isochrones suggest an age of $\sim$ 2.0 Gyr for this cluster,
leaving ESO 121 as the only known LMC cluster with an age between 
approximately 3 and 13 Gyr.
} 
\label{fig:n1718_cmd}
\end{center}
\end{figure}

While only three of the stars observed in NGC 1718 appear to be cluster
members, these stars are, on average, 0.3 dex more metal poor than all but
one of the other stars observed in this field. As mentioned in \S
\ref{sec:properties}, this causes NGC 1718 to occupy an interesting
position in the LMC's age/metallicity relation; its metallicity is
comparable to that of ESO 121, which seems to be the only cluster residing
in the LMC having an age between $\sim$ 3 and 13 Gyr (Da Costa 2002). The
cluster CMD resulting from our aperture photometry is not well populated
around the MSTO, so we have used archival HST/WFPC2 data (GO-5475) to
create a cluster CMD reaching below the MSTO.  The images were reduced
using the procedure outlined by Sarajedini (1998). In summary, all
detected stars on the Planetary Camera CCD were photometered in the F450W
and F555W filters using a small aperture.  These were then corrected to a
0.5 arcsec radius, adjusted for the exposure time, and transformed to the
standard system using the equations from Holtzman et al. (1995). In
Fig.~\ref{fig:n1718_cmd} we present the CMD of NGC 1718 with isochrones
from Girardi et al.~(2002) overplotted; the isochrones have [Fe/H]
$\approx$ -0.7, close to our measured cluster value of -0.8 dex, and ages
ranging from 1.3 to 2.5 Gyr.  This figure suggests that NGC 1718 has an
age of roughly 2.0 Gyr, making it an intermediate age cluster and leaving
ESO 121 as still the only cluster known to occupy the LMC's cluster `age
gap.' However, the existence of an intermediate age cluster at this low
metallicity is intriguing as it indicates that some pockets of unenriched
material must have remained intact even though most of the gas which
formed the intermediate age clusters was well mixed.

\subsection{NGC 1846}

Given the sloped appearance of the RC and the width of the RGB, NGC 1846
is suffering from differential reddening, making it difficult to
accurately measure the true location of the cluster RC as well as
$V-V_{HB}$ for target stars.  To address this problem, we make no 
adjustments to the instrumental magnitudes, but we measure the median 
magnitude of the entire differentially reddened RC, effectively 
measuring the RC at the mean reddening of the cluster.  Since the amount 
of extinction suffered by the RGB stars should be scattered about the mean 
reddening, this approach will `smooth over' the differential reddening, 
allowing us to accurately measure the cluster metallicity.  We note that 
this method will increase the scatter in [Fe/H] for cluster members; as 
such, we have relaxed the metallicity cut in our member selection method 
to include all stars moving at the radial velocity of the cluster.  For 
reference, if $V-V_{HB}$ for any given star is off by $\pm$ 0.2 mag (we 
estimate that the differential reddening is 0.4 magnitudes in V), the 
effect on [Fe/H] for that star is roughly $\pm$ 0.05.

\subsection{NGC 1861}

This cluster is listed as SWB Type IVB, suggesting an age range of
0.4$-$0.8 Gyr (Bica et al.~1996), which is roughly the age when the RC
first forms ($\sim$ 0.5 Gyr; Girardi \& Salaris, 2001).  Plotting a CMD of
stars within the apparent cluster radius reveals what appears to be a
fairly young MSTO in addition to no obvious cluster RC or RGB. Therefore,
we assume that NGC 1861 is a young cluster and all observed RGB stars are
actually part of an older field population.

\clearpage



\clearpage

\clearpage

\clearpage

\clearpage

\clearpage

\clearpage

\clearpage

\clearpage

\clearpage

\clearpage

\clearpage

\clearpage

\clearpage

\clearpage

\begin{deluxetable}{lccccccccccc}
\tabletypesize{\scriptsize}
\tablecaption{LMC Target Cluster Information\label{table:info1}}
\tablewidth{0pt}
\tablehead{
\colhead{Cluster}&
\colhead{Alternate}& 
\colhead{RA\tablenotemark{a}}&
\colhead{Dec\tablenotemark{a}}&
\colhead{Diameter\tablenotemark{a}}&
\colhead{V mag\tablenotemark{b}}&
\colhead{SWB}&
\colhead{Exposure}
\\
& \colhead{name}& (J2000)& (J2000)& (arcmin)& & 
\colhead{Type\tablenotemark{b}}& \colhead{Time (s)}
}
\startdata
SL 4      & LW 4  &4$^h$32$^m$38$^s$& -72$\degr$20$\arcmin$27$\arcsec$& 
1.7\tablenotemark{c}& 14.2& VI\tablenotemark{c}& 2$\times$300\\
Reticulum & ESO 118-SC31  &4 36 11& -58 51 40& 4.7& 14.25& VII& 
2$\times$600 
\\
NGC 1651  & SL 7, LW 12  &4 37 33& -70 35 08& 2.7& 12.28& V& 
2$\times$300\\
NGC 1652  & SL 10, LW 14  &4 38 23& -68 40 22& 1.5& 13.13& VI& 
300\\
NGC 1841  & ESO 4-SC15  &4 45 23& -83 59 49& 0.9& 11.43& VII& 
500\\
SL 41     & LW 64  &4 47 30& -72 35 18& 1.4& 14.14& V& 
2$\times$600\\
SL 61     & LW 79  &4 50 45& -75 32 00& 2.3& 13.99& VI& 
2$\times$300\\
NGC 1718  & SL 65  &4 52 25& -67 03 06& 1.8& 12.25& VI& 500\\
NGC 1751  & SL 89  &4 54 12& -69 48 23& 1.5& 11.73& VI& 
2$\times$300\\
NGC 1846  & SL 243  &5 07 35& -67 27 31& 3.8& 11.31& VI& 600\\
NGC 1861  & SL 286  &5 10 21& -70 46 38& 1.5& 13.16& IVB& 600\\
SL 396    & LW 187  &5 19 36& -73 06 40& 1.3& 13.56& VI& 
2$\times$300\\
NGC 1942  & SL 445, LW 203  &5 24 43& -63 56 24& 1.9& 13.46& VI& 
2$\times$300\\
NGC 2019  & SL 554  &5 31 57& -70 09 34& 1.5& 10.86& VII& 600\\
Hodge 4   & SL 556, LW 237  &5 32 25& -64 44 12& 2.5& 13.33& V& 
500\\
Hodge 3   & SL 569  &5 33 20& -68 08 08& 1.8& 13.42& VI& 500\\
IC 2146   & SL 632, LW 258  &5 37 46& -74 47 00& 3.3& 12.41& V& 
500\\
SL 663    & LW 273  &5 42 29& -65 21 48& 0.8\tablenotemark{c}& 
13.8 & V\tablenotemark{c}& 600\\
NGC 2121  & SL 725, LW 303  &5 48 12& -71 28 52& 2.4& 12.37& VI& 
600\\
NGC 2173  & SL 807, LW 348  &5 57 58& -72 58 41& 2.6& 11.88& VI& 
600\\
NGC 2155  & SL 803, LW 347  &5 58 33& -65 28 35& 2.4& 12.60& VI& 
600\\
NGC 2162  & SL 814, LW 351  &6 00 30& -63 43 19& 3.0& 12.70& V& 
2$\times$300\\
NGC 2203  & SL 836, LW 380  &6 04 43& -75 26 18& 3.2& 11.29& VI& 
500\\
NGC 2193  & SL 839, LW 387  &6 06 18& -65 05 57& 1.7& 13.42& V& 
600\\
NGC 2213  & SL 857, LW 419  &6 10 42& -71 31 44& 2.1& 12.38& V& 
2$\times$300\\
Hodge 11  & SL 868, LW 437  &6 14 22& -69 50 54& 2.7& 11.93& VII& 
600\\
SL 869    & LW 441  &6 14 41& -69 48 07& 1.6\tablenotemark{c}& 
15.0 & VI\tablenotemark{c}& 600\\
NGC 2231  & SL 884, LW 466  &6 20 43& -67 31 07& 2.1& 13.20& V& 
500\\
NGC 2257  & SL 895, LW 481  &6 30 13& -64 19 29& 4.0& 12.62& VII& 
600\\

\enddata
\tablenotetext{a}{From Bica et al.~(1999).}
\tablenotetext{b}{From Bica et al.~(1996) unless noted.}
\tablenotetext{c}{Estimated via comparisons with target clusters.}

\end{deluxetable}

\begin{deluxetable}{lccc}
\tablecaption{CaT line and continuum bandpasses\label{table:bandpass}}
\tablewidth{0pt}
\tablehead{
\colhead{Feature}&
\colhead{Line Bandpass (\AA)}&
\colhead{Blue Continuum (\AA)}&
\colhead{Red Continuum (\AA)} 
}
\startdata
\ion{Ca}{2} $\lambda$8498 &8490 $-$ 8506 &8474 $-$ 8489 &8521 $-$ 8531 \\
\ion{Ca}{2} $\lambda$8542 &8532 $-$ 8552 &8521 $-$ 8531 &8555 $-$ 8595 \\
\ion{Ca}{2} $\lambda$8662 &8653 $-$ 8671 &8626 $-$ 8650 &8695 $-$ 8725 \\
\enddata

\end{deluxetable}


\begin{deluxetable}{lccccccccccc}
\tablewidth{0pt}
\tabletypesize{\scriptsize}
\tablecaption{Positions and Measured Values for All Cluster Members
\label{table:all_stars} }
\tablehead{
\colhead{ID} &
\colhead{RA} &
\colhead{Dec} &
\colhead{RV} &
\colhead{$\sigma_{RV}$} &
\colhead{$V-V_{HB}$} &
\colhead{$\Sigma W$} &
\colhead{$\sigma_{\Sigma W}$} &
\\
& (J2000) & (J2000) & (km s$^{-1}$) & (km s$^{-1}$) & (mag) & (\AA) & 
(\AA)
}
\startdata
& & & \it{SL 4} & & & & &\\
 6&  4$^h$32$^m$41$\fs$65&  -72$\degr$20$\arcmin$59$\farcs$1&  237.4&
    6.7&  -0.89&  7.65&  0.23& \\
 7&  4 32 40.46&  -72 20 50.3&  221.6&   6.8&  -0.64&  7.09&  0.22& \\
 9&  4 32 36.78&  -72 20 32.2&  234.5&   6.8&  -0.70&  6.76&  0.31& \\
11&  4 32 41.01&  -72 20 12.3&  221.0&   6.9&  -1.07&  7.59&  0.16& \\
12&  4 32 39.18&  -72 20 00.4&  221.3&   6.9&  -0.03&  7.27&  0.36& \\
& & & & & & & &\\
& & & \it{Reticulum} & & & & &\\
 2&  4 36 13.17&  -58 52 49.6&  250.9&   6.8&  -1.44&  5.37&  0.12& \\
 3&  4 36 14.49&  -58 52 40.1&  254.3&   6.8&  -0.81&  4.34&  0.19& \\
 5&  4 36 18.63&  -58 52 19.6&  244.1&   7.3&   0.78&  3.60&  0.48& \\
 6&  4 36 17.19&  -58 52 09.6&  249.6&   7.1&  -0.06&  3.53&  0.33& \\
 8&  4 36 10.73&  -58 51 46.6&  245.8&   7.1&   0.36&  3.19&  0.46& \\
 9&  4 36 06.80&  -58 51 30.3&  251.3&   6.9&  -1.03&  4.33&  0.17& \\
11&  4 36 07.52&  -58 51 11.5&  252.8&   7.1&  -0.53&  4.00&  0.16& \\
12&  4 36 08.07&  -58 51 00.6&  250.0&   7.0&  -0.52&  4.06&  0.24& \\
13&  4 36 04.51&  -58 50 50.6&  237.0&   6.7&  -0.55&  4.47&  0.22& \\
14&  4 36 01.47&  -58 50 33.3&  243.1&   6.8&   0.23&  3.88&  0.47& \\
15&  4 36 09.60&  -58 50 23.9&  239.9&   7.1&  -0.33&  3.66&  0.31& \\
16&  4 36 02.14&  -58 50 12.6&  252.6&   7.3&   0.07&  4.32&  0.39& \\
17&  4 36 14.10&  -58 49 57.0&  245.8&   6.8&  -1.16&  4.98&  0.12& \\
& & & & & & & &\\
& & & \it{NGC 1651} & & & & &\\
 6&  4 37 34.49&  -70 35 50.2&  235.4&   6.9&  -0.39&  6.82&  0.25& \\
 8&  4 37 33.25&  -70 35 31.7&  235.5&   6.7&   0.44&  6.73&  0.79& \\
10&  4 37 37.35&  -70 35 13.9&  223.2&   6.8&   0.27&  6.18&  0.46& \\
11&  4 37 34.63&  -70 35 05.5&  233.3&   6.8&  -1.14&  7.88&  0.41& \\
12&  4 37 34.76&  -70 35 02.6&  225.4&   6.8&  -0.87&  7.29&  0.17& \\
13&  4 37 29.94&  -70 34 53.8&  223.1&   6.8&  -0.01&  6.43&  0.35& \\
15&  4 37 31.06&  -70 34 35.2&  237.2&   6.8&  -1.07&  7.74&  0.14& \\
16&  4 37 31.29&  -70 34 25.0&  219.2&   6.8&  -2.26&  8.43&  0.10& \\
17&  4 37 29.58&  -70 34 16.0&  221.7&   7.1&  -0.23&  6.87&  0.26& \\
& & & & & & & &\\
& & & \it{NGC 1652} & & & & &\\
 5&  4 38 21.50&  -68 40 55.4&  271.3&   6.8&   0.02&  6.85&  0.34& \\
 6&  4 38 23.75&  -68 40 47.1&  281.3&   6.9&  -0.63&  7.68&  0.26& \\
 7&  4 38 23.43&  -68 40 39.6&  276.7&   6.8&   0.08&  6.50&  0.52& \\
 8&  4 38 19.31&  -68 40 30.6&  272.7&   6.8&  -0.55&  7.34&  0.19& \\
 9&  4 38 19.84&  -68 40 20.9&  276.3&   6.8&  -1.59&  8.50&  0.14& \\
10&  4 38 22.03&  -68 40 11.1&  273.9&   6.9&  -0.70&  7.41&  0.22& \\
12&  4 38 20.33&  -68 39 46.5&  277.9&   6.9&   0.17&  6.48&  0.48& \\
& & & & & & & &\\
& & & \it{NGC 1841} & & & & &\\
 1&  4 44 05.58&  -84 01 24.0&  212.0&   6.8&  -1.60&  3.51&  0.16& \\
 2&  4 44 52.68&  -84 01 14.3&  205.2&   6.9&  -1.79&  3.88&  0.12& \\
 4&  4 45 27.22&  -84 00 53.5&  211.8&   7.0&  -1.18&  3.72&  0.16& \\
 6&  4 45 41.54&  -84 00 33.7&  212.3&   6.9&  -2.11&  4.25&  0.10& \\
 8&  4 44 42.94&  -84 00 09.1&  212.8&   6.9&  -0.88&  3.41&  0.20& \\
 9&  4 45 19.87&  -83 59 59.0&  210.4&   6.8&  -1.49&  3.55&  0.18& \\
10&  4 45 33.89&  -83 59 50.4&  204.3&   6.8&  -2.25&  4.24&  0.12& \\
11&  4 45 10.36&  -83 59 40.7&  204.6&   6.8&  -1.56&  3.80&  0.15& \\
12&  4 45 15.53&  -83 59 29.8&  214.3&   7.1&  -1.27&  3.62&  0.23& \\
13&  4 45 43.84&  -83 59 17.7&  209.8&   6.9&  -0.93&  3.46&  0.22& \\
14&  4 45 12.82&  -83 59 02.5&  207.4&   6.9&  -0.70&  3.30&  0.30& \\
15&  4 45 28.38&  -83 58 48.4&  211.3&   7.4&  -1.52&  3.71&  0.12& \\
16&  4 45 20.51&  -83 58 38.4&  215.8&   6.8&  -2.60&  4.64&  0.10& \\
17&  4 46 22.26&  -83 58 28.7&  212.1&   6.9&  -1.04&  3.18&  0.20& \\
19&  4 45 55.01&  -83 58 00.9&  208.9&   7.0&  -1.29&  3.60&  0.19& \\
20&  4 45 22.59&  -83 57 45.5&  211.6&   6.8&  -2.46&  4.00&  0.11& \\
& & & & & & & &\\
& & & \it{SL 41} & & & & &\\
 4&  4 47 30.78&  -72 35 55.6&  223.8&   7.1&  -0.42&  7.29&  0.31& \\
 5&  4 47 29.32&  -72 35 42.7&  229.0&   6.9&  -0.13&  7.39&  0.20& \\
 6&  4 47 28.06&  -72 35 32.0&  228.5&   6.8&  -0.85&  7.40&  0.15& \\
 7&  4 47 31.86&  -72 35 22.1&  230.5&   6.7&  -1.89&  8.20&  0.11& \\
 8&  4 47 32.39&  -72 35 14.1&  230.5&   6.8&  -2.33&  8.75&  0.10& \\
 9&  4 47 30.92&  -72 35 06.0&  233.4&   7.0&  -0.21&  7.06&  0.22& \\
& & & & & & & &\\
& & & \it{SL 61} & & & & &\\
 4&  4 50 49.97&  -75 32 45.0&  224.0&   6.8&  -0.58&  7.58&  0.27& \\
 8&  4 50 44.33&  -75 32 07.1&  226.1&   6.8&  -1.07&  7.50&  0.22& \\
 9&  4 50 48.31&  -75 31 59.2&  222.3&   6.6&  -1.84&  8.55&  0.18& \\
10&  4 50 34.50&  -75 31 48.6&  215.8&   7.0&   0.20&  7.32&  0.62& \\
11&  4 50 44.07&  -75 31 41.2&  227.8&   6.8&  -1.71&  8.32&  0.12& \\
12&  4 50 50.74&  -75 31 31.6&  225.5&   6.7&  -0.15&  7.07&  0.43& \\
13&  4 50 42.63&  -75 31 22.9&  211.3&   7.3&   0.14&  7.69&  0.61& \\
14&  4 50 44.32&  -75 31 14.1&  222.8&   6.8&  -2.41&  9.30&  0.11& \\
& & & & & & & &\\
& & & \it{NGC 1718} & & & & &\\
 8&  4 52 27.20&  -67 03 25.7&  277.5&   6.7&  -1.30&  7.03&  0.25& \\
10&  4 52 27.67&  -67 03 08.6&  275.3&   6.7&  -1.82&  7.34&  0.11& \\
13&  4 52 20.40&  -67 02 38.5&  282.6&   6.8&  -0.59&  6.26&  0.25& \\
& & & & & & & &\\
& & & \it{NGC 1751} & & & & &\\
 5&  4 54 19.46&  -69 49 18.9&  242.8&   6.7&  -2.21&  8.55&  0.11& \\
 6&  4 54 01.54&  -69 49 09.0&  237.7&   6.9&  -1.26&  7.25&  0.21& \\
 8&  4 54 07.81&  -69 48 48.1&  249.7&   6.7&  -2.99&  9.48&  0.14& \\
10&  4 54 08.30&  -69 48 20.1&  252.1&   6.7&  -1.63&  8.30&  0.15& \\
11&  4 54 13.67&  -69 48 07.2&  244.8&   6.7&  -2.06&  8.70&  0.11& \\
14&  4 54 15.30&  -69 47 39.1&  245.4&   6.8&  -1.56&  8.09&  0.17& \\
& & & & & & & &\\
& & & \it{NGC 1846} & & & & &\\
 2&  5 07 39.69&  -67 28 59.4&  233.2&   6.7&  -1.81&  8.37&  0.11& \\
 4&  5 07 32.76&  -67 28 31.0&  236.2&   6.7&  -1.39&  8.20&  0.14& \\
 5&  5 07 39.03&  -67 28 22.3&  231.0&   6.8&  -1.52&  8.03&  0.12& \\
 6&  5 07 35.07&  -67 28 13.2&  240.1&   6.7&  -0.05&  6.92&  0.29& \\
 8&  5 07 33.80&  -67 27 56.4&  241.1&   7.0&  -0.12&  6.64&  0.42& \\
 9&  5 07 33.87&  -67 27 54.6&  229.5&   7.0&  -1.45&  7.94&  0.11& \\
10&  5 07 36.86&  -67 27 45.0&  234.3&   6.8&  -2.49&  9.28&  0.15& \\
11&  5 07 39.26&  -67 27 35.1&  234.8&   6.7&  -1.98&  8.14&  0.12& \\
12&  5 07 30.12&  -67 27 26.5&  238.0&   6.8&  -1.77&  8.43&  0.10& \\
13&  5 07 40.15&  -67 27 14.6&  235.2&   8.4&  -0.63&  7.08&  0.46& \\
14&  5 07 35.98&  -67 27 06.2&  235.3&   6.7&  -1.51&  8.00&  0.14& \\
15&  5 07 38.81&  -67 26 57.7&  230.5&   6.7&  -2.10&  8.49&  0.11& \\
16&  5 07 32.46&  -67 26 48.5&  232.9&   6.7&  -1.55&  7.70&  0.12& \\
17&  5 07 33.62&  -67 26 40.4&  236.8&   6.8&  -1.98&  8.51&  0.12& \\
18&  5 07 34.04&  -67 26 31.8&  231.4&   6.8&  -0.37&  6.63&  0.22& \\
20&  5 07 43.43&  -67 26 12.6&  235.9&   6.8&  -0.35&  7.16&  0.20& \\
21&  5 07 33.58&  -67 26 02.6&  242.4&   6.9&  -0.33&  6.34&  0.29& \\
& & & & & & & &\\
& & & \it{SL 396} & & & & &\\
 8&  5 19 38.32&  -73 06 57.8&  228.4&   6.9&  -0.65&  7.17&  0.18& \\
 9&  5 19 35.96&  -73 06 43.7&  224.3&   6.7&  -1.48&  8.14&  0.11& \\
10&  5 19 38.67&  -73 06 37.3&  221.7&   6.8&  -0.91&  7.93&  0.17& \\
11&  5 19 38.97&  -73 06 25.7&  226.1&   6.8&  -1.83&  8.89&  0.13& \\
13&  5 19 41.07&  -73 06 03.8&  225.6&   6.8&   0.00&  6.92&  0.35& \\
& & & & & & & &\\
& & & \it{NGC 1942} & & & & &\\
 6&  5 24 44.67&  -63 56 47.2&  286.8&   6.8&  -1.65&  8.37&  0.13& \\
 7&  5 24 45.95&  -63 56 38.8&  298.4&   6.8&   0.00&  6.75&  0.24& \\
 8&  5 24 43.59&  -63 56 30.0&  286.6&   6.8&  -1.68&  7.63&  0.43& \\
 9&  5 24 46.05&  -63 56 20.1&  289.6&   6.8&  -2.43&  9.04&  0.12& \\
10&  5 24 46.56&  -63 56 11.8&  301.0&   6.8&  -0.42&  7.21&  0.20& \\
11&  5 24 46.80&  -63 56 01.5&  299.5&   6.6&  -0.02&  6.72&  0.33& \\
12&  5 24 41.90&  -63 55 51.6&  287.8&   6.7&  -0.83&  7.54&  0.15& \\
13&  5 24 41.84&  -63 55 37.9&  300.2&   7.0&  -0.06&  6.33&  0.28& \\
& & & & & & & &\\
& & & \it{NGC 2019} & & & & &\\
 8&  5 32 00.40&  -70 10 19.1&  281.3&   6.7&  -2.59&  6.32&  0.09& \\
10&  5 32 02.96&  -70 10 01.4&  287.2&   6.8&  -2.41&  6.83&  0.13& \\
11&  5 31 59.36&  -70 09 51.9&  283.6&   6.7&  -2.60&  6.59&  0.17& \\
12&  5 32 03.01&  -70 09 43.9&  274.1&   6.8&  -1.73&  5.70&  0.14& \\
13&  5 31 55.94&  -70 09 35.0&  276.8&   6.8&  -1.61&  5.39&  0.14& \\
& & & & & & & &\\
& & & \it{Hodge 4} & & & & &\\
 6&  5 32 24.81&  -64 44 56.5&  317.0&   6.8&   0.17&  6.02&  0.29& \\
 7&  5 32 25.96&  -64 44 47.5&  301.8&   6.8&   0.07&  6.67&  0.30& \\
 8&  5 32 28.14&  -64 44 38.0&  309.6&   6.8&  -0.80&  7.75&  0.14& \\
 9&  5 32 24.60&  -64 44 30.0&  310.9&   6.8&  -0.82&  7.08&  0.16& \\
10&  5 32 27.95&  -64 44 18.4&  312.4&   6.8&  -0.35&  7.11&  0.21& \\
12&  5 32 26.11&  -64 43 59.7&  315.1&   7.0&  -0.16&  6.37&  0.22& \\
13&  5 32 25.97&  -64 43 47.3&  308.7&   6.8&  -1.73&  8.41&  0.12& \\
& & & & & & & &\\
& & & \it{Hodge 3} & & & & &\\
 5&  5 33 14.78&  -68 09 50.9&  277.0&   6.8&  -2.16&  8.30&  0.10& \\
 7&  5 33 11.87&  -68 09 29.3&  280.1&   6.8&  -0.24&  7.75&  0.33& \\
 8&  5 33 19.42&  -68 09 20.3&  279.7&   6.8&  -0.97&  7.68&  0.20& \\
10&  5 33 22.35&  -68 09 10.0&  275.8&   7.2&  -1.54&  8.67&  0.15& \\
11&  5 33 16.19&  -68 08 59.8&  277.1&   6.8&  -1.36&  8.32&  0.15& \\
12&  5 33 19.61&  -68 08 50.1&  277.9&   6.8&  -2.12&  9.25&  0.12& \\
13&  5 33 20.49&  -68 08 35.5&  274.4&   6.8&  -1.60&  8.54&  0.15& \\
& & & & & & & &\\
& & & \it{IC 2146} & & & & &\\
 2&  5 37 42.00&  -74 48 31.0&  224.1&   6.7&  -0.35&  7.04&  0.50& \\
 3&  5 37 37.67&  -74 48 20.0&  225.4&   6.8&  -1.57&  8.37&  0.15& \\
 4&  5 37 48.70&  -74 48 11.0&  225.3&   6.8&  -1.88&  8.67&  0.15& \\
 5&  5 37 43.94&  -74 48 00.2&  224.1&   6.8&  -2.44&  9.24&  0.12& \\
 7&  5 37 38.53&  -74 47 40.3&  227.6&   6.7&  -1.55&  8.17&  0.15& \\
 8&  5 37 53.30&  -74 47 31.6&  227.2&   6.9&  -0.15&  7.25&  0.49& \\
 9&  5 37 42.15&  -74 47 23.4&  223.5&   6.8&  -1.13&  7.60&  0.17& \\
10&  5 37 43.16&  -74 47 13.3&  224.4&   6.8&  -0.65&  7.71&  0.54& \\
11&  5 37 51.92&  -74 47 03.4&  229.5&   6.7&  -1.46&  8.36&  0.13& \\
12&  5 37 44.66&  -74 46 54.6&  227.3&   6.8&  -0.61&  7.52&  0.30& \\
13&  5 37 45.72&  -74 46 46.5&  222.6&   6.8&  -1.31&  7.85&  0.16& \\
14&  5 37 47.30&  -74 46 36.7&  226.4&   6.8&  -1.15&  7.44&  0.21& \\
15&  5 37 40.09&  -74 46 26.6&  231.5&   6.8&  -0.10&  7.34&  0.40& \\
16&  5 37 40.38&  -74 46 15.2&  232.3&   6.8&  -2.49&  9.22&  0.12& \\
17&  5 37 50.48&  -74 46 06.6&  226.0&   6.6&  -0.14&  6.70&  0.56& \\
18&  5 38 03.62&  -74 45 54.4&  225.1&   6.8&  -2.22&  8.80&  0.11& \\
19&  5 38 03.11&  -74 45 45.1&  225.7&   6.9&  -1.05&  7.60&  0.17& \\
21&  5 37 41.58&  -74 45 16.8&  224.9&   6.8&  -1.99&  8.58&  0.11& \\
& & & & & & & &\\
& & & \it{SL 663} & & & & &\\
 5&  5 42 29.19&  -65 22 35.1&  299.1&   6.7&  -1.89&  8.57&  0.10& \\
 8&  5 42 30.17&  -65 21 59.6&  297.3&   6.8&  -1.17&  7.44&  0.13& \\
 9&  5 42 29.59&  -65 21 47.3&  301.6&   6.8&  -1.69&  8.43&  0.15& \\
10&  5 42 26.31&  -65 21 37.7&  303.3&   7.0&  -0.06&  6.28&  0.30& \\
11&  5 42 29.92&  -65 21 29.5&  299.8&   6.7&  -0.07&  6.73&  0.27& \\
12&  5 42 29.07&  -65 21 19.4&  300.2&   6.8&  -1.08&  7.24&  0.14& \\
13&  5 42 28.41&  -65 21 05.4&  299.2&   6.8&   0.03&  6.90&  0.36& \\
14&  5 42 29.73&  -65 20 53.8&  311.1&   6.8&  -0.17&  6.59&  0.32& \\
& & & & & & & &\\
& & & \it{NGC 2121} & & & & &\\
 2&  5 48 10.29&  -71 30 16.8&  228.4&   6.7&  -2.04&  8.60&  0.11& \\
 7&  5 48 07.09&  -71 29 24.4&  236.5&   6.7&  -1.30&  7.25&  0.19& \\
 8&  5 48 11.85&  -71 29 10.9&  234.7&   6.8&  -1.98&  8.36&  0.10& \\
 9&  5 48 06.03&  -71 29 01.3&  233.9&   6.8&  -1.90&  8.51&  0.11& \\
10&  5 48 22.29&  -71 28 46.2&  235.4&   6.7&  -1.45&  7.96&  0.13& \\
11&  5 48 12.54&  -71 28 38.0&  231.4&   6.8&  -0.74&  7.56&  0.18& \\
12&  5 48 09.70&  -71 28 29.8&  239.7&   6.8&  -1.65&  7.83&  0.12& \\
14&  5 48 15.15&  -71 28 12.3&  236.2&   6.8&  -1.70&  8.12&  0.11& \\
15&  5 48 15.71&  -71 28 01.6&  229.7&   6.8&  -2.14&  8.51&  0.12& \\
16&  5 47 57.44&  -71 27 53.2&  227.6&   6.9&   0.03&  6.37&  0.23& \\
17&  5 48 25.44&  -71 27 45.2&  227.9&   6.8&  -1.62&  7.81&  0.11& \\
20&  5 48 13.58&  -71 27 11.7&  229.0&   6.8&  -1.87&  8.15&  0.12& \\
& & & & & & & &\\
& & & \it{NGC 2173} & & & & &\\
 6&  5 57 54.09&  -72 59 11.1&  238.5&   6.8&  -1.37&  7.76&  0.16& \\
 8&  5 57 55.77&  -72 58 52.4&  237.4&   6.8&  -2.33&  8.91&  0.12& \\
 9&  5 57 52.62&  -72 58 42.2&  235.6&   6.8&  -2.21&  8.67&  0.13& \\
10&  5 57 57.94&  -72 58 29.8&  238.6&   6.8&  -1.87&  8.24&  0.25& \\
11&  5 57 59.18&  -72 58 20.8&  235.4&   6.8&  -1.04&  7.79&  0.19& \\
14&  5 57 58.02&  -72 57 53.9&  239.3&   6.8&  -2.88&  9.41&  0.11& \\
& & & & & & & &\\
& & & \it{NGC 2155} & & & & &\\
 3&  5 58 36.33&  -65 29 39.0&  310.8&   6.8&  -1.23&  7.94&  0.14& \\
 5&  5 58 37.83&  -65 29 13.2&  311.7&   6.7&  -0.83&  7.11&  0.16& \\
 6&  5 58 32.85&  -65 29 03.4&  314.2&   6.8&  -2.19&  8.84&  0.12& \\
 7&  5 58 32.27&  -65 28 54.3&  308.8&   6.8&  -0.14&  7.04&  0.22& \\
 9&  5 58 28.86&  -65 28 38.4&  305.9&   6.8&  -1.72&  7.98&  0.10& \\
10&  5 58 35.94&  -65 28 29.3&  311.0&   6.8&  -1.80&  8.32&  0.12& \\
14&  5 58 35.67&  -65 27 41.6&  301.5&   6.7&  -0.07&  6.21&  0.23& \\
& & & & & & & &\\
& & & \it{NGC 2162} & & & & &\\
 7&  6 00 27.09&  -63 43 37.6&  323.0&   6.8&  -1.24&  7.49&  0.14& \\
 8&  6 00 28.17&  -63 43 28.7&  318.1&   6.9&  -1.20&  7.40&  0.25& \\
 9&  6 00 32.04&  -63 43 18.0&  320.4&   6.8&  -2.52&  8.78&  0.13& \\
10&  6 00 29.47&  -63 43 02.1&  335.6&   6.8&  -1.88&  8.39&  0.13& \\
11&  6 00 27.80&  -63 42 51.7&  315.9&   6.8&   0.07&  7.52&  0.49& \\
& & & & & & & &\\
& & & \it{NGC 2203} & & & & &\\
 7&  6 04 48.16&  -75 26 48.1&  244.6&   6.7&  -1.81&  8.18&  0.12& \\
 8&  6 04 40.98&  -75 26 39.8&  251.2&   6.8&  -1.78&  8.15&  0.12& \\
 9&  6 04 35.08&  -75 26 29.2&  246.2&   6.7&  -1.80&  8.60&  0.14& \\
10&  6 04 41.31&  -75 26 21.2&  249.7&   6.8&  -1.71&  8.39&  0.12& \\
11&  6 04 33.33&  -75 26 13.1&  240.1&   6.8&  -2.07&  8.72&  0.79& \\
12&  6 04 45.42&  -75 26 04.0&  247.9&   6.8&  -1.68&  8.10&  0.12& \\
13&  6 04 49.22&  -75 25 55.6&  247.7&   6.8&  -2.84&  9.55&  0.14& \\
14&  6 04 43.24&  -75 25 45.6&  242.2&   6.8&  -0.74&  7.71&  0.19& \\
16&  6 04 50.33&  -75 25 24.2&  239.8&   6.8&  -1.25&  7.71&  0.14& \\
& & & & & & & &\\
& & & \it{NGC 2193} & & & & &\\
 6&  6 06 20.10&  -65 06 08.0&  288.4&   6.8&  -2.08&  8.71&  0.10& \\
 7&  6 06 20.74&  -65 05 58.6&  287.9&   6.7&  -1.55&  7.78&  0.11& \\
 8&  6 06 18.03&  -65 05 48.0&  298.8&   6.8&  -1.63&  8.20&  0.15& \\
 9&  6 06 15.56&  -65 05 32.3&  290.0&   6.7&  -0.18&  6.58&  0.25& \\
10&  6 06 16.58&  -65 05 24.1&  291.2&   6.8&  -0.08&  6.97&  0.31& \\
& & & & & & & &\\
& & & \it{NGC 2213} & & & & &\\
 7&  6 10 41.62&  -71 32 19.0&  238.6&   6.7&  -1.51&  7.84&  0.11& \\
 8&  6 10 45.00&  -71 32 00.1&  247.1&   6.8&  -2.18&  8.66&  0.11& \\
 9&  6 10 45.68&  -71 31 47.1&  244.1&   6.8&  -1.53&  8.05&  0.13& \\
10&  6 10 43.45&  -71 31 36.8&  240.2&   6.8&  -0.53&  7.08&  0.23& \\
11&  6 10 39.14&  -71 31 33.7&  242.7&   7.1&  -1.95&  8.20&  0.11& \\
12&  6 10 44.25&  -71 31 09.4&  243.4&   6.8&  -0.51&  6.74&  0.20& \\
& & & & & & & &\\
& & & \it{Hodge 11} & & & & &\\
 3&  6 14 31.61&  -69 49 36.5&  245.2&   6.8&  -1.47&  4.15&  0.12& \\
 4&  6 14 30.51&  -69 49 51.1&  251.3&   6.9&  -0.71&  3.36&  0.17& \\
 7&  6 14 21.85&  -69 49 57.6&  246.6&   6.8&  -2.01&  4.16&  0.10& \\
 8&  6 14 24.80&  -69 50 16.6&  247.5&   7.0&  -0.87&  3.18&  0.23& \\
 9&  6 14 24.55&  -69 50 28.4&  243.5&   6.8&  -2.87&  5.57&  0.17& \\
10&  6 14 21.99&  -69 50 33.4&  240.8&   6.8&  -3.03&  5.84&  0.10& \\
11&  6 14 16.88&  -69 50 28.0&  241.9&   7.0&  -0.38&  3.74&  0.27& \\
12&  6 14 17.20&  -69 50 43.8&  241.6&   6.8&  -2.13&  4.54&  0.08& \\
13&  6 14 17.88&  -69 50 54.3&  244.6&   6.8&  -2.03&  4.67&  0.08& \\
14&  6 14 22.57&  -69 51 18.8&  242.0&   7.3&  -1.14&  4.10&  0.18& \\
15&  6 14 07.08&  -69 50 46.5&  250.9&   7.2&  -0.38&  3.71&  0.24& \\
18&  6 14 08.43&  -69 51 22.9&  245.2&   8.0&  -0.14&  2.82&  0.26& \\
& & & & & & & &\\
& & & \it{SL 869} & & & & &\\
 7&  6 14 39.08&  -69 47 30.6&  262.5&   6.8&  -1.37&  7.89&  0.14& \\
 9&  6 14 40.73&  -69 47 59.6&  256.3&   6.8&  -1.17&  7.93&  0.13& \\
11&  6 14 44.28&  -69 48 46.4&  256.4&   6.7&  -1.27&  8.21&  0.14& \\
& & & & & & & &\\
& & & \it{NGC 2231} & & & & &\\
 3&  6 20 41.59&  -67 31 45.3&  274.5&   6.8&  -1.09&  7.23&  0.14& \\
 4&  6 20 43.41&  -67 31 34.7&  272.6&   6.8&  -0.32&  6.75&  0.22& \\
 5&  6 20 38.01&  -67 31 18.3&  274.5&   6.7&  -2.21&  8.73&  0.11& \\
 6&  6 20 45.55&  -67 31 06.2&  279.4&   6.7&  -0.53&  7.40&  0.28& \\
 7&  6 20 45.24&  -67 30 57.3&  282.7&   6.8&  -1.50&  7.94&  0.12& \\
 8&  6 20 44.94&  -67 30 46.1&  278.6&   6.7&  -1.22&  7.39&  0.20& \\
 9&  6 20 48.86&  -67 30 36.2&  285.1&   6.8&  -0.51&  7.27&  0.18& \\
10&  6 20 40.00&  -67 30 24.2&  274.0&   6.8&  -1.94&  8.27&  0.12& \\
11&  6 20 40.63&  -67 30 02.9&  277.2&   6.7&  -0.90&  7.39&  0.18& \\
& & & & & & & &\\
& & & \it{NGC 2257} & & & & &\\
 3&  6 30 09.09&  -64 20 33.3&  302.9&   7.0&  -0.37&  3.94&  0.16& \\
 4&  6 30 14.36&  -64 20 23.0&  301.1&   6.9&  -2.30&  5.48&  0.08& \\
 5&  6 30 14.94&  -64 20 10.0&  300.6&   6.9&  -1.19&  4.76&  0.12& \\
 6&  6 30 14.45&  -64 20 01.9&  303.5&   6.9&  -0.99&  4.73&  0.13& \\
 7&  6 30 07.92&  -64 19 53.0&  299.9&   7.0&  -0.41&  4.13&  0.16& \\
 8&  6 30 12.16&  -64 19 43.5&  308.0&   6.8&  -1.14&  4.63&  0.18& \\
 9&  6 30 14.78&  -64 19 32.7&  296.6&   7.0&  -1.90&  4.88&  0.08& \\
10&  6 30 10.73&  -64 19 24.3&  303.4&   7.0&  -0.21&  4.02&  0.27& \\
11&  6 30 09.22&  -64 19 15.4&  297.6&   6.9&  -0.80&  4.58&  0.16& \\
12&  6 30 06.66&  -64 19 06.8&  297.5&   6.9&  -0.36&  4.10&  0.21& \\
13&  6 30 14.97&  -64 18 58.5&  298.9&   6.8&  -1.85&  5.17&  0.09& \\
14&  6 30 11.03&  -64 18 46.6&  307.1&   7.0&  -1.17&  4.74&  0.14& \\
15&  6 30 16.85&  -64 18 33.8&  302.8&   6.9&  -0.80&  4.08&  0.14& \\
17&  6 30 27.48&  -64 18 07.5&  300.7&   6.9&  -0.82&  4.19&  0.15& \\
18&  6 30 16.26&  -64 17 53.8&  304.7&   7.1&  -0.20&  4.12&  0.22& \\
19&  6 30 11.25&  -64 17 44.1&  299.7&   7.3&  -0.21&  4.21&  0.24&
\enddata
\end{deluxetable}

\begin{deluxetable}{lccccccccccc}
\tabletypesize{\scriptsize}
\tablecaption{Derived LMC Cluster Properties\label{table:properties}}
\tablewidth{0pt}
\tablehead{
\colhead{Cluster} & \colhead{n stars}& \colhead{RV}& 
\colhead{$\sigma_{\overline{RV}}$} & \colhead{[Fe/H]} & 
\colhead{$\sigma_{\overline{[Fe/H]}}$} & 
\\
\colhead{Name} & & \colhead{(km s$^{-1}$)} & \colhead{(km s$^{-1}$)} & 
\colhead{(dex)} & \colhead{(dex)}  
}
\startdata
SL 4      &  5&  227.1&   3.6&  -0.51&  0.06  \\
Reticulum & 13&  247.5&   1.5&  -1.57&  0.03  \\
NGC 1651  &  9&  228.2&   2.3&  -0.53&  0.03  \\
NGC 1652  &  7&  275.7&   1.3&  -0.46&  0.04  \\
NGC 1841  & 16&  210.3&   0.9&  -2.02&  0.02  \\
SL 41     &  6&  229.3&   1.3&  -0.44&  0.03  \\
SL 61     &  8&  221.9&   2.0&  -0.35&  0.04  \\
NGC 1718  &  3&  278.4&   2.2&  -0.80&  0.03  \\
NGC 1751  &  6&  245.4&   2.1&  -0.44&  0.05  \\
NGC 1846  & 17&  235.2&   0.9&  -0.49&  0.03  \\
NGC 1861  & ... & ... & ... & ... & ...  \\
SL 396    &  5&  225.2&   1.1&  -0.39&  0.05  \\
NGC 1942  &  8&  293.7&   2.3&  -0.50&  0.04  \\
NGC 2019  &  5&  280.6&   2.3&  -1.31&  0.05  \\
Hodge 4   &  7&  310.8&   1.9&  -0.55&  0.06  \\
Hodge 3   &  7&  277.4&   0.8&  -0.32&  0.05  \\
IC 2146   & 18&  226.3&   0.6&  -0.41&  0.02  \\
SL 663    &  8&  301.4&   1.5&  -0.54&  0.05  \\
NGC 2121  & 12&  232.5&   1.2&  -0.50&  0.03  \\
NGC 2173  &  6&  237.4&   0.7&  -0.42&  0.03  \\
NGC 2155  &  7&  309.1&   1.6&  -0.50&  0.05  \\
NGC 2162  &  5&  322.6&   3.5&  -0.46&  0.07  \\
NGC 2203  &  9&  245.5&   1.4&  -0.41&  0.03  \\
NGC 2193  &  5&  291.2&   2.0&  -0.49&  0.05  \\
NGC 2213  &  6&  242.7&   1.2&  -0.52&  0.04  \\
Hodge 11  & 12&  245.1&   1.0&  -1.84&  0.04  \\
SL 869    &  3&  258.4&   2.1&  -0.40&  0.04  \\
NGC 2231  &  9&  277.6&   1.4&  -0.52&  0.03  \\
NGC 2257  & 16&  301.6&   0.8&  -1.59&  0.02  \\

\enddata

\end{deluxetable}

\begin{deluxetable}{lccccccccccc}
\tabletypesize{\scriptsize}
\tablecaption{Published LMC Cluster Metallicities\label{table:other_data}}
\tablewidth{0pt}
\tablehead{
\colhead{Cluster} & \colhead{[Fe/H]} & 
\colhead{[Fe/H]\tablenotemark{a}} &
\colhead{[Fe/H]\tablenotemark{b}}&
\colhead{[Fe/H]} \\
\colhead{name} & (this work) & CaT & CaT & Hi-Res 
}
\startdata
SL 4      & $-0.51$& ... & ... & ... \\
Reticulum & $-1.57$& $-1.71$\tablenotemark{c} & $-1.44$\tablenotemark{d} 
(9) & ... \\
NGC 1651  & $-0.53$& $-0.37$ & $-0.41$ (0.5) & ... \\
NGC 1652  & $-0.46$& $-0.45$ & $-0.46$ (2) & ... \\
NGC 1841  & $-2.02$& $-2.11$\tablenotemark{c} & $-1.83$\tablenotemark{d} 
(8) & $-2.07$\tablenotemark{f}\\
SL 41     & $-0.44$& ... & ... & ... \\
SL 61     & $-0.35$& $-0.50$ & $-0.49$ (1) & ... \\
NGC 1718  & $-0.80$& ... & ... & ... \\
NGC 1751  & $-0.44$& $-0.18$ & $-0.31$ (0.5) & ... \\
NGC 1846  & $-0.49$& $-0.70$ & $-0.62$ (1) & ... \\
SL 396    & $-0.39$& ... & ... & ... \\
NGC 1942  & $-0.50$& $+0.16$ & $-0.14$ (1) & ... \\
NGC 2019  & $-1.31$& $-1.81$ & $-1.53$ (1) & $-1.24$\tablenotemark{e} (3)\\
Hodge 4   & $-0.55$& $-0.15$ & $-0.29$ (1) & ... \\
Hodge 3   & $-0.32$& ... & ... & ... \\
IC 2146   & $-0.41$& $-0.40$ & $-0.43$ (2) & ... \\
SL 663    & $-0.54$& ... & ... & ... \\
NGC 2121  & $-0.50$& $-0.61$ & $-0.56$ (1.5) & ... \\
NGC 2173  & $-0.42$& $-0.24$ & $-0.34$ (1) & ... \\
NGC 2155  & $-0.50$& $-0.55$ & $-0.52$ (2.5) & ... \\
NGC 2162  & $-0.46$& $-0.23$ & $-0.33$ (2) & ... \\
NGC 2203  & $-0.41$& $-0.52$ & $-0.51$ (2) & ... \\
NGC 2193  & $-0.49$& ... & ... & ... \\
NGC 2213  & $-0.52$& $-0.01$ & $-0.22$ (1) & ... \\
Hodge 11  & $-1.84$& $-2.06$ & $-1.78$ (2) & $-2.13$\tablenotemark{e} (2)\\
SL 869    & $-0.40$& ... & ... & ... \\
NGC 2231  & $-0.52$& $-0.67$ & $-0.60$ (1.5) & ... \\
NGC 2257  & $-1.59$& ... & ... & $-1.86$\tablenotemark{f} \\
\enddata



\tablenotetext{a}{From OSSH, unless otherwise noted.}
\tablenotetext{b}{From OSSH, unless otherwise noted, converted onto our
system using Eq.~\ref{eq:ossh_conversion}.}
\tablenotetext{c}{From Suntzeff et al.~(1992).}
\tablenotetext{d}{From Suntzeff et al.~(1992), converted onto our system 
using Eq.~\ref{eq:ossh_conversion}.}
\tablenotetext{e}{From Johnson et al.~(2006)}
\tablenotetext{f}{From Hill (2004)}

\end{deluxetable}

\begin{deluxetable}{lccccccccccc}
\tablecaption{Metallicities of Young and 
Intermediate Age Stellar Populations\label{table:metal_compare}}
\tablewidth{0pt}
\tablehead{
\colhead{Population} & \colhead{Age Estimate} & 
\colhead{[Fe/H]} &
\colhead{$\sigma_{[Fe/H]}$}&
\colhead{Reference} \\
&(Myr)&&&
}
\startdata

B dwarfs & $<$20 & $-0.31$ & 0.04 & Rolleston et al.~(2002)\\
Cepheid variables & 10$-$60 & $-0.34$ & 0.15 & Luck et al.~(1998)\\
Young RGB stars & 200$-$1000 & $-0.45$ & 0.10 & Smith et al.~(2002)\\
Intermediate age clusters & 1000$-$3000 & $-0.48$ & 0.09 & This paper\\
Intermediate age clusters & 1000$-$3000 & $-0.48$ & 0.17 & OSSH\\
Bar RGB stars, metal rich & 1000$-$5000 & $-0.37$ & 0.15 & Cole et 
al.~(2005)\\
Bar RGB stars, metal poor & $\gea$5000 & $-1.08$ & 0.47 & Cole et 
al.~(2005)\\

\enddata
\end{deluxetable}


\end{document}